\documentclass[english,notitlepage]{revtex4-1}
\usepackage[T1]{fontenc}
\usepackage[latin9]{inputenc}
\usepackage{babel}
\usepackage{refstyle}
\usepackage{amsmath}
\usepackage{amssymb}
\usepackage{graphicx}
\usepackage[unicode=true,pdfusetitle,
 bookmarks=true,bookmarksnumbered=false,bookmarksopen=false,
 breaklinks=false,pdfborder={0 0 1},backref=false,colorlinks=false]
 {hyperref}

\makeatletter

\AtBeginDocument{\providecommand\subsecref[1]{\ref{subsec:#1}}}
\providecommand{\tabularnewline}{\\}
\RS@ifundefined{subsecref}
  {\newref{subsec}{name = \RSsectxt}}
  {}
\RS@ifundefined{thmref}
  {\def\RSthmtxt{theorem~}\newref{thm}{name = \RSthmtxt}}
  {}
\RS@ifundefined{lemref}
  {\def\RSlemtxt{lemma~}\newref{lem}{name = \RSlemtxt}}
  {}

\makeatother

\begin{document}

\title{Network reconstruction from infection cascades}

\author{Alfredo Braunstein}
\email{alfredo.braunstein@polito.it}

\affiliation{DISAT, Politecnico di Torino, Corso Duca Degli Abruzzi 24, 10129
Torino}

\affiliation{Human Genetics Foundation, Via Nizza 52, 10124 Torino}

\affiliation{Collegio Carlo Alberto, Via Real Collegio 1, Moncalieri}

\author{Alessandro Ingrosso}
\email{ai2367@columbia.edu}

\affiliation{Center for Theoretical Neuroscience, Columbia University, New York,
USA}

\author{Anna Paola Muntoni}
\email{anna.muntoni@polito.it}

\affiliation{DISAT, Politecnico di Torino, Corso Duca Degli Abruzzi 24, 10129
Torino}
\begin{abstract}
Accessing the network through which a propagation dynamics diffuse
is essential for understanding and controlling it. In a few cases,
such information is available through direct experiments or thanks
to the very nature of propagation data. In a majority of cases however,
available information about the network is indirect and comes from
partial observations of the dynamics, rendering the network reconstruction
a fundamental inverse problem. Here we show that it is possible to
reconstruct the whole structure of an interaction network and to simultaneously
infer the complete time course of activation spreading, relying just
on single epoch (i.e. snapshot) or time-scattered observations of
a small number of activity cascades. The method that we present is
built on a Belief Propagation approximation, that has shown impressive
accuracy in a wide variety of relevant cases, and is able to infer
interactions in presence of incomplete time-series data by providing
a detailed modeling of the posterior distribution of trajectories
conditioned to the observations. Furthermore, we show by experiments
that the information content of full cascades is relatively smaller
than that of sparse observations or single snapshots.
\end{abstract}
\maketitle
Much effort has been devoted recently to the inverse problem of reconstructing
the topology of a network from time series of a dynamical process
acting on it. The methods proposed so far in the literature heavily
rely on complete knowledge of the dynamical trajectories of some spreading
process. In certain cases, when information about the time-series
of the process is available, the problem can be, and has been, cast
into relatively simple terms, since a sequence of time-consecutive
states of a pair of nodes gives direct information about the potential
interaction between them. In many cases, however, the set of available
observations is much sparser, possibly on a much slower timescale
than that of the dynamics, and often skipping the initial stages of
the propagation which would give precious information about the initial
condition. In particular, in an observation consisting on a single
snapshot of the system there is no \emph{direct} information about
the interaction of nodes, as evidence of interaction indeed comes
from variation of the state of nodes in time.

Let us take the example of second messenger cascades in a cell, and
suppose the experimenter has access to the expression profile of a
huge number of proteins in different cascades. Monitoring the exact
time course of the concentration of each protein is currently challenging,
if not unfeasible: one observes a concerted up and down-regulation
of a big number of proteins, which naturally follow from a complex
time course in a network of reciprocal protein-protein interactions.
One is confronted with a similar information shortage in the context
of epidemic spreading in a network of individuals: there is no information
about who was the first one to contract the disease, and little is
known about the underlying networks of contacts between individuals,
which may even be dynamically changing over time.

Even though direct experimental data about contact networks in diverse
contexts is being collected at a fast rate \citep{Salathe21122010,isella_whats_2011,rocha_information_2010},
there are some strong experimental and technical limitations to this
collection, sometimes due to privacy protection regulations or concerns.
However, knowledge of propagation networks would have a large list
of benefits. First, it may allow to understand the propagation process
better, including finding entry-points (e.g. the so called \emph{index
case} or \emph{patient zero} in epidemiological jargon) of an ongoing
epidemic. Second, it may allow to devise strategies to control the
process in various ways, for example hindering the propagation (e.g.
targeted vaccination) or favoring it (e.g. in the context of maximizing
information diffusion on social networks, in viral or targeted advertising,
etc). In this respect, a number of computational studies have introduced
optimization methods based on message-passing that address the problem
of containing \citep{altarelli2014containing}or maximizing spreading
\citep{altarelli_optimizing_2013,lokhov2016optimal}.

Recently, several approaches have been proposed for the problem of
deducing the propagation network from time-series, based on a naive
Bayes approach \citep{lokhov2015efficient} with efficient computations
based on dynamic message-passing equations \citep{karrer_message_2010,lokhov_inferring_2013},
compressed sensing schemes \citep{shen2014reconstructing}, genetic
\citep{wan_inferring_2014} and dynamic programming algorithms \citep{wang_identifying_2016},
tensor decomposition \citep{yang_characterizing_2017} or on Monte
Carlo sampling \citep{li_reconstruction_2017}. These methods all
share the need for observations at consecutive epochs.

Despite this recent progress, in most contexts the available observations
of each cascade are sparse, noisy and discontinuous in time. In such
situations, none of the methods proposed in the literature can be
applied. One example is the problem of inferring functional contacts
in signaling pathways, in which interacting proteins generate cascades
of phosphorilation which eventually transmit signals from the cell
membrane to the nucleus. Observations come in general from gene expression
data, and the network to be inferred is a subnetwork of a large-scale
protein-protein interaction network (PPI), also known as \textit{interactome}.
Although several experimental and computational approaches are able
to identify candidate links of these networks, they lack in distinguishing
false positive from true positive links \citep{BORK2004292,VALENCIA2002368,Yu2006}
that seems to be a challenging task. Social science and epidemiology
offer another interesting domain of application, as one generally
tries to infer the network of social contacts from a limited amount
of sparse and noisy observations of some propagation histories.

Here we present a Bayesian technique that allows to uncover the complete
functional structure (including its topology and parameters) of a
network from a limited amount of single snapshots of the state of
the network cascades. Starting from a functional parametrization of
the posterior probability distribution of propagation trajectories,
our technique builds on a Message Passing procedure that allows to
compute, and then maximize, the likelihood of a given network structure.
This computation can be performed efficiently thanks to Belief Propagation
(BP), which is proven to be exact for tree graphs and has been successfully
used in a variety of problems in general graphs with loops. Upon convergence,
the parameters allow both to identify the network and the sources
of the infection for each cascade with great accuracy. The method
is effective on progressive propagation models like susceptible-infected
(SI), susceptible-infected-removed (SIR), independent cascades (IC)
and variants, including models with hidden variables (e.g. representing
latency times). We called this method Gradient Ascent Belief Propagation
(GABP)

Although the proposed inference machinery is very general, we focus
on the well known Susceptible-Infected-Recovered (SIR)\citep{kermack_contribution_1927}
model, which describes those diseases in which infected individuals
become immune to future infections after recovery (such as measles,
rubella, chicken pox and generic influenza). More generally, SIR constitutes
a good model for the spreading of rumor and information over a network
or for the interaction dynamics among proteins.

Our minimal model of activity propagation in a network is very simple:
if a node $i$ is active (Infected) at time $t$, it has a finite
probability $\lambda_{ij}$ to activate (or infect) any of its neighbors
$j$, which will in turn be active at time $t+1$. An active node
will recover in each time-step with a (generally site-dependent) recovery
probability $\mu_{i}$. Once recovered, individuals do not get sick
anymore, and will not be able to infect other nodes. This will result
in a propagation throughout the network, that we call a cascade.

\begin{figure}
\centering{}\includegraphics[width=0.6\columnwidth]{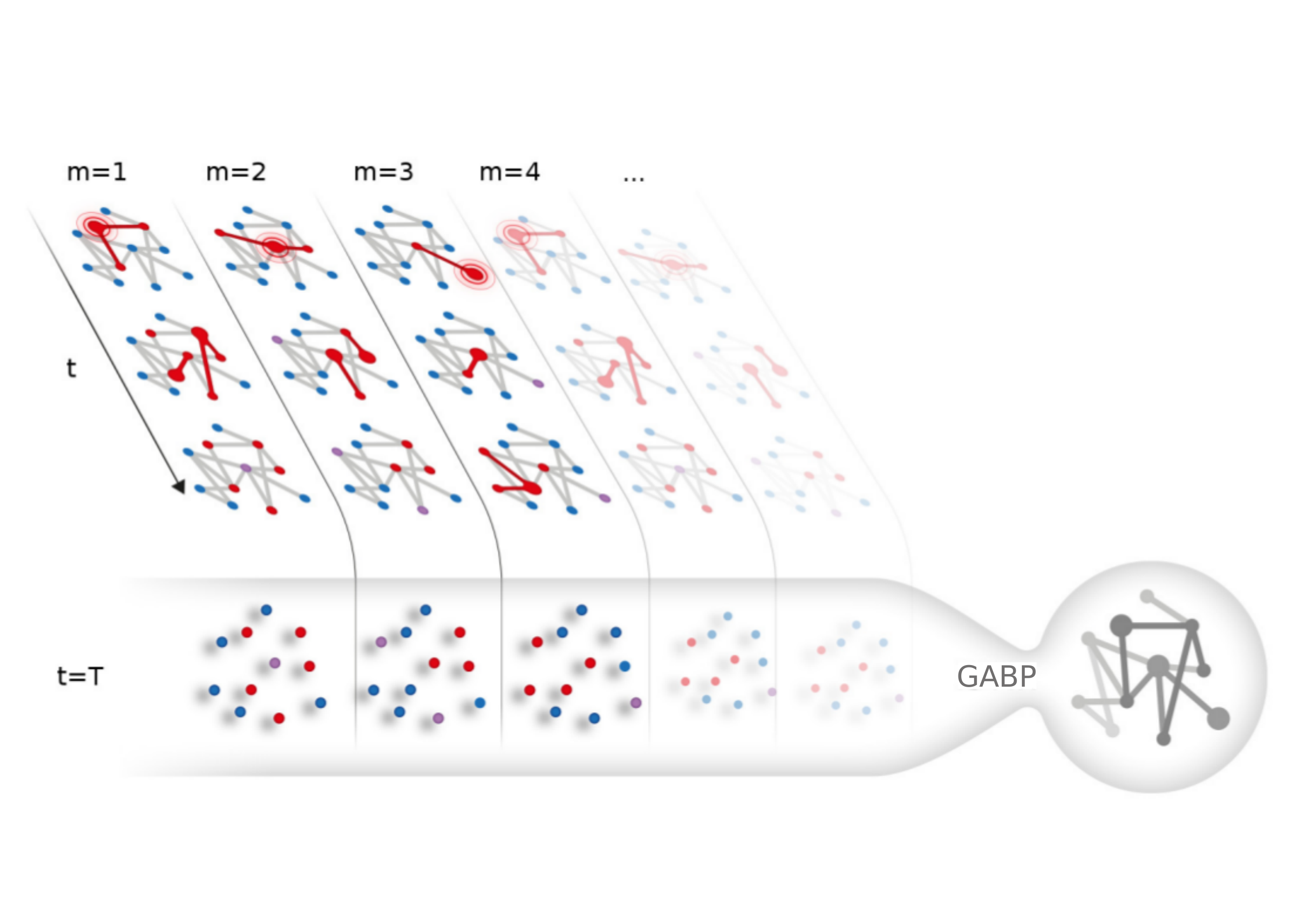}\caption{\label{fig:idnr}Cartoon representation of the network reconstruction
problem: $M$ independent cascades starting from different sources
(highlighted in the first frame of each vertical stripe) are represented,
with time flowing downward. Infected nodes are red, Susceptible nodes
are blue and Recovered nodes are purple. The GABP algorithm is provided
a set of $M$ snapshots taken $T$ time steps after the cascade onset:
the goal is to reconstruct the functional interactions in the network
$G$ as well as to identify the source of each cascade.}
\end{figure}

Let us then suppose that a number $M$ of \textit{independent} realizations
(or cascades) of the SIR dynamics can be observed. With independent
cascades we mean that the realization of each stochastic process (in
terms of infection and recovery times) do not affect the dynamics
of other cascades. Identical realizations (same infected and recovered
nodes, same infection and recovery times) are obviously strongly correlated
but this does not limit the applicability of the method as far as
each process is independent of one another. In the prototypical situation,
the complete history of the propagation is not available: all we can
observe is a number of \textquotedblleft frozen\textquotedblright{}
snapshots of a wave-front of the activity at a given time $T$, when
all the states of the nodes in the network can be assessed to a reasonable
extent of accuracy. Our aim is to identify the hidden network structure
and the set of transmission probabilities for each link. Fig. \ref{fig:idnr}
shows a cartoon representation of the problem.

\section*{Results}

\subsection*{A static formulation of the dynamical process}

Reconstructing the unknown connectivity structure of the network is
inevitably coupled to that of tracing back in time the entire history
of the spreading process for each cascade $m$, $m\in\left\{ 1,...,M\right\} $,
which in turn results in the identification of the sources of diffusion.
Our approach builds on computing a joint posterior probability distribution
over all cascades that are compatible with the observations, and then
maximizing the likelihood of interaction parameters of the network
at the same time. To set our notation, let us consider a weighted
undirected graph $G=\left(V,E,\Lambda,\Omega\right)$ with a number
$|G|$ of nodes, where $\Lambda=\left\{ \lambda_{ij}\right\} _{ij\in E}$
play the role of edge-dependent infection probabilities in a SIR stochastic
model, and that is also equipped with a set $\Omega=\left\{ \mu_{i}\right\} _{i\in V}$
of site-dependent recovery probabilities. For directed graphs, we
allow parameters $\lambda_{ij}\neq\lambda_{ji}$. Focusing, for the
moment, on a single cascade, at any point in time each node $i$ will
be in one of three possible states: susceptible $(S)$, infected $(I)$,
and recovered/removed $(R)$. The state of node $i$ at time $t$
in each cascade $m$ is represented by a variable $x_{i}\left(t\right)\in\{S,I,R\}$,
with $t$ in some discrete set. At each time step (e.g. a day) of
the stochastic dynamics, an infected node $i$ can first spread the
disease to each susceptible neighbor $j$ with given probability $\lambda_{ij}$,
then recover with probability $\mu_{i}$. Each cascade is defined
by the set of vectors $\mathbf{x}^{m}\left(t\right)$, with $m$ labeling
the cascade, and we assume that for each cascade the initial state
$\mathbf{x}^{m}\left(0\right)$ is composed of just one Infected node
$i_{0}^{m}$, with all the other nodes in the network in the being
in the Susceptible state. We will assume that we have access to the
state of the nodes in the networks only $T^{m}=T$ steps after the
initiation of each cascade.

Let us consider a node $i$ which gets infected at its infection time
$t_{i}$: since it has a finite probability to pass the disease to
a neighbor $j$ in each time step, this results in a stochastic transmission
delay $s_{ij}$. In addition, the individual $i$ recovers at time
$t_{i}+g_{i}$, with $g_{i}$ a stochastic recovery delay. Owing to
the irreversibility of the spreading process, each cascade is fully
specified by the quantities $\left\{ t_{i},g_{i}\right\} _{i\in V}$
and $\left\{ s_{ij}\right\} _{(i,j)\in E}$ for each node and each
link in the network. It is then possible to construct a simple static
graphical model representation of the dynamical process for each cascade
on the grounds of the following simple observation: the time at which
a given node $i$ gets infected only depends on the infection times
of its neighbors $j$, and the infection delays of these nodes. Infection
times $t_{i}>0$ are related by the deterministic equations 
\begin{equation}
t_{i}=1+\min_{j\in\partial i}\left\{ t_{j}+s_{ji}\right\} \label{eq:dynamical}
\end{equation}
which are a set of $|G|$ constraints encoding the infection dynamics,
involving only local quantities at each node. Once the initial condition
$\boldsymbol{x}\left(0\right)$ and stochastic quantities $s_{ij}$
and $g_{i}$ are thrown independently from their own distributions,
the infection times are given deterministically by virtue of equation
(\ref{eq:dynamical}).

This observation was exploited in a series of works \citep{altarelli_optimizing_2013,altarelli_bayesian_2014,altarelli_patient-zero_2014}
to develop a fully Bayesian method for approximating the whole probability
distribution of the time evolution of the system, conditioned on some
observations, and was originally used to identify the origin of the
epidemic outbreak in SIR and similar models. The method is built on
a Belief Propagation approximation (see Methods), which is exact on
tree graphs and has proven successful in general networks with loops.

What if the underlying network is unknown, and so are the epidemic
parameters $\left\{ \lambda_{ij},\mu_{i}\right\} $? In a Maximum
Likelihood approach, one needs to define the quantity $\mathcal{P}\left(\left\{ \mathbf{x}^{m}\left(T\right)\right\} |\left\{ \lambda_{ij}\right\} ,\left\{ \mu_{i}\right\} \right),$
namely the likelihood of epidemic parameters with respect to observations,
and then be able to maximize over the relevant parameters. Note that
in a fully Bayesian framework, incorporating \emph{a priori} information
on the network topology or epidemic parameters is straightforward:
it would lead to add a log-prior term $f_{\lambda,\mu}=\log\mathcal{P_{\lambda}}\left(\left\{ \lambda_{ij}\right\} \right)+\log\mathcal{P_{\mu}}\left(\left\{ \mu_{i}\right\} \right)$
to the log-likelihood to obtain a log-posterior. The log-likelihood
of the parameters coincides with the so-called free-entropy of the
system $\mathcal{L}\left(\left\{ \lambda_{ij}\right\} ,\left\{ \mu_{i}\right\} \right)=\log\mathcal{P}\left(\left\{ \mathbf{x}^{m}\left(T\right)\right\} |\left\{ \lambda_{ij}\right\} ,\left\{ \mu_{i}\right\} \right)=-f\left(\left\{ \lambda_{ij}\right\} ,\left\{ \mu_{i}\right\} \right)$,
which can be computed, consistently with the BP approximation, employing
the Bethe decomposition (see Methods).

The BP method for the (cavity) marginal distributions of infection
times can be then interleaved with simple log-likelihood climbing
steps in a Gradient Ascent (GA) scheme, leading to a unique set of
equations that are solved by iteration. In this setting, the computation
of the gradient of the log-likelihood relies only on local updates
involving the BP cavity messages. Ultimately all the information has
to be processed locally at each node. That, in addition to other simplifications,
entails a huge reduction of computational time, making the analysis
of large-scale networks feasible efficiently (see Methods). One starts
from a flat assignment of the parameters, and the initial fully connected
network gets progressively pruned by means of the GA updates, eventually
leading to a reconstructed network strongly resembling the real one.

\subsection*{Reconstructing random networks}

We start by investigating three basic random network structures, namely
Random Regular (RR), Erdos-Renyi (ER) and Barabasi-Albert (BA) scale-free
networks: an impressive level of accuracy may be reached with a small
number $M$ of observations. In a RR network, each node is connected
at random with a fixed number of neighbors in the networks, whereas
in the ER graph the number of neighbors is Poisson distributed. Scale-free
networks, on the other hand, possess a power law degree distribution,
and are known to capture some key ingredients of many real networks
encountered in practical applications (for a review, see \citep{barabasi_scalefree}).

As a first step, a random graph is constructed, and a set of $M$
cascades are simulated, each one being an independent realization
of the stochastic SIR process with a random initial source $i_{0}^{m}$.
GABP is then run until the parameters $\lambda_{ij}$ and $\mu_{i}$
reach a stable value. Since the goal of the inference is two-fold,
we use two different measures of the inference performance. For each
cascade $m$, the nodes in the network are ranked in decreasing order
with respect of the estimated probability of being the origin of the
observed epidemic: the ability to identify the sources of the spreading
is easily quantified by the rank of $i_{0}^{m}$, namely the position
of $i_{0}^{m}$ in the ordered list.

On the other hand, a simple method for quantifying the accuracy of
network reconstruction is the \emph{Receiver Operating Characteristic}
(\emph{ROC}) curve, namely a plot of the \emph{true positive rate}
against the \emph{false positive rate} in a binary classification
problem. Constructing the \emph{ROC} curve in the present case is
very easy: the inferred values of $\lambda_{ij}$ are ranked in decreasing
order, and one step upward in the \emph{ROC} is taken if the link
is present in the original graph (true positive) or one step rightward
if the link is absent (false positive). The area under the \emph{ROC}
curve is a good indication of the discrimination ability: areas close
to one signal a good discrimination between true links and non existent
links. The reconstruction performances are compared to those of an
empirical correlations based method. For each possible couple of nodes
we compute, at the time of the observation $T$, the probability of
having an edge $\left(i,\,j\right)$ as the mutual information (MI)
between node $i$ and $j$; details of the calculations are reported
in \subsecref{Mutual-information}. As for the case of parameters
$\lambda_{ij}$, we construct \textit{ROC} curves and we compute \textit{ROC}
areas from the set of correlation measures $m_{ij}$.

\begin{figure*}
\noindent \begin{centering}
\includegraphics[width=1\textwidth]{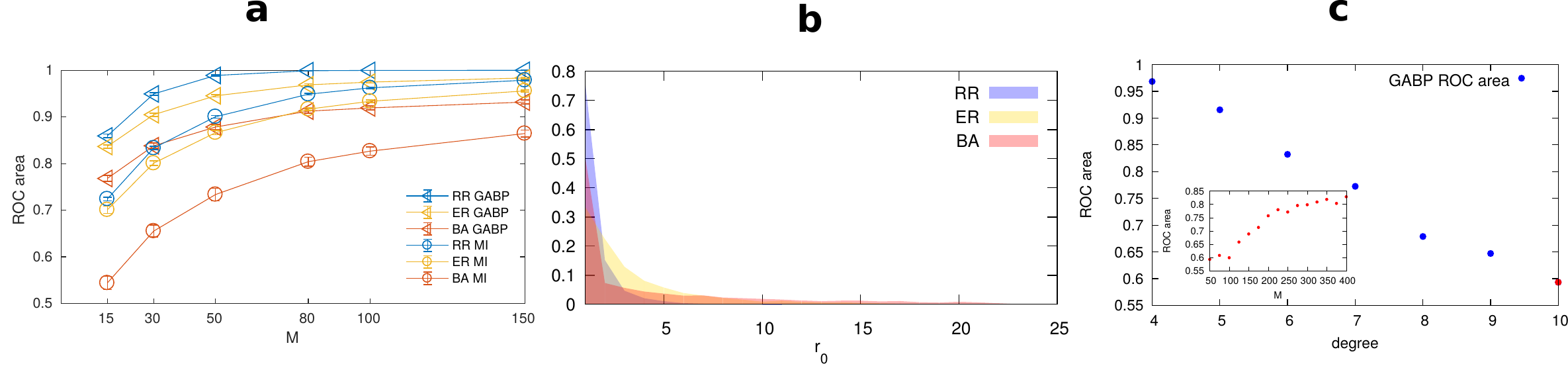}
\par\end{centering}
\noindent \centering{}\caption{\label{fig:random_graphs_reconstruction}\textbf{a}: Reconstruction
accuracy in three types of random networks using GABP and MI. Each
curve is an average over $30$ random instances of the area under
the \emph{ROC} curve, as a function of the number of observed cascades
$M$ at time $T=5$. Epidemic parameters, $\lambda_{ij}=0.6$ and
$\mu_{i}=0.4$, are the same for all the three type of networks. The
size of the network is $|G|=50$. Blue curve (RR): Random Regular
graphs with degree $d=4$; red curve (BA): Barabasi-Albert (scale-free)
networks with average degree $d_{av}=4$; yellow curve (ER): Erdos-Renyi
graphs with average degree $d_{av}=4$. Triangular and circular marks
show GABP and MI results, respectively. \textbf{b}: Identification
of initial spreaders. Each filled curve is the histogram of the rank
of the true patient zero $i_{0}^{m}$ at $M=150$ for the three types
of network. Histograms refer to $30$ random instances, thus considering
a total of $30*150=4500$ independent cascades. \textbf{c}: Reconstruction
accuracy versus connectivity. The blue curve is the area under the
\emph{ROC} curve in different instances of Random Regular graphs of
size $|G|=50$ with increasing degree $d$. In each case, $M=50$
cascades are observed at time $T=7$. Recovery rate is fixed to$\mu_{i}=0.4$,
$\lambda_{ij}$ is scaled down as degree increases in order to keep
the size of epidemics roughly constant.\emph{ Inset}: area under the
ROC curve as a function of the number of observed cascades $M$ in
a random regular graph with degree $d=10$ (corresponding to the red
point at the end of the blue curve in the main plot).}
\end{figure*}

We report in Fig. \ref{fig:random_graphs_reconstruction} (\textbf{a})
a systematic investigation of the reconstruction performances of GABP
and MI in the three types of random networks with an increasing number
of cascades $M.$ The parameters of the infection are $\lambda=0.6$
and $\mu=0.4$ for all the experiments. For all values of $M$ GABP
outperforms the MI method as the\textit{ ROC} areas associated with
the GABP predictions are notably greater then the one obtained from
MI. In the case of BA graphs, we notice smaller values of the ROC
areas because, for these values of the parameters of the SIR dynamics,
we observe huge epidemies in which at time $T$ almost all nodes are
infected or recovered. This efficient spreading is caused by the presence
of hubs that easily infect a good portion of the network in one time-step.
In this regime and even for large value of $M$ there is not sufficient
information to fully recover the true links of the graphs.

The ability to identify the sources of spreading (patient zero) is
easily quantified by the rank $r_{0}^{(m)}$ of the true patient zero
$i_{0}^{m}$ in each of the $M$ cascade: if $M$ is high enough so
that enough information is conveyed on the underlying network structure,
GABP is able to successfully identify most of the true initial spreaders
in each cascade. This can be seen in Fig. \ref{fig:random_graphs_reconstruction}
(\textbf{b}), that shows the distribution of $r_{0}^{(m)}$ for a
value of $M=150$ in the three types of random networks considered
here, which is fairly concentrated on low values of $r_{0}^{(m)}$.

The reconstruction performance is expected to be substantially related
to the density of the network. This can be investigated by systematically
varying the degree of connectivity of a network, as it is shown in
Fig. \ref{fig:random_graphs_reconstruction} (\textbf{c}), where the
performance of GABP is assessed in a RR graph of size $|G|=50$ with
an increasing connectivity degree $d$, from $d=4$ to $d=10$. The
accurate reconstruction of denser networks requires, consequently,
a larger number of cascades $M$.

\begin{figure*}
\centering{}\includegraphics[width=1\textwidth]{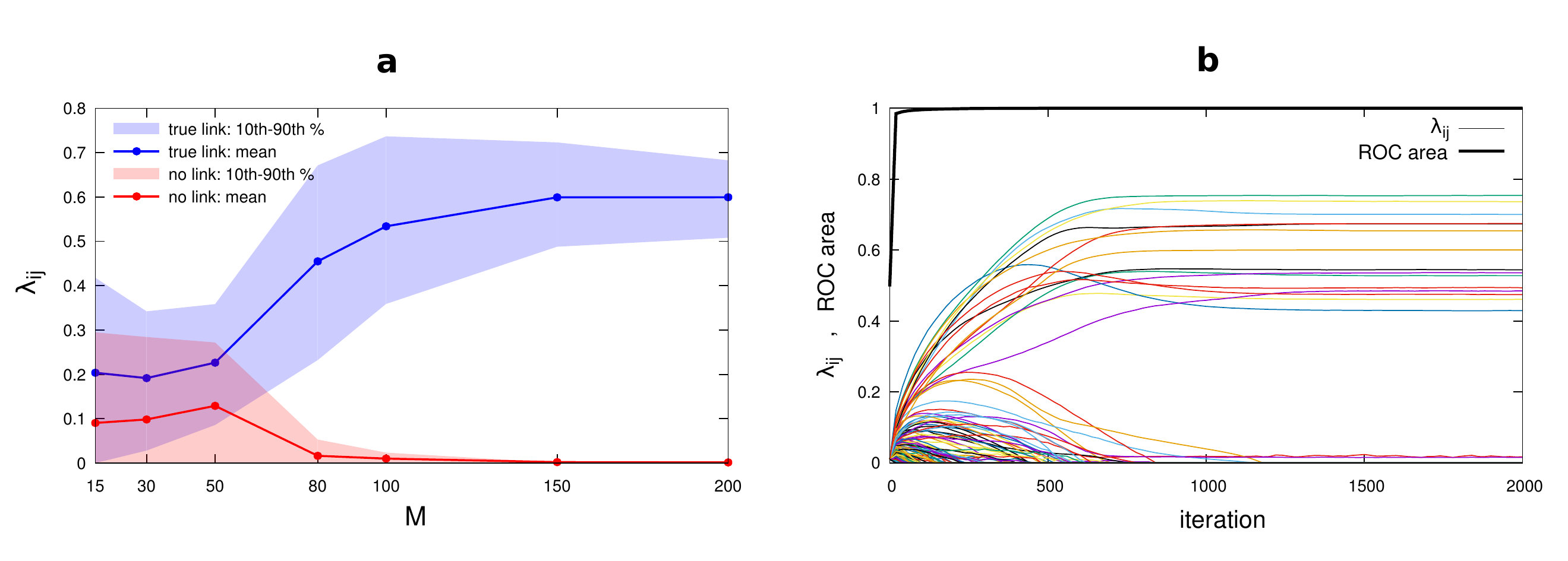}\caption{\label{fig:lambda_dynamics}GABP rapidly identifies true links. \textbf{a}:
average value of $\lambda_{ij}$ for true links (blue) versus non
existent ones (red) as a function of the number of observed cascades
in a Random Regular graph with size $|G|=50$, $\lambda_{ij}=0.6$,
$\mu_{i}=0.4$ and $d=4$; shaded areas correspond to the intervals
between the $10$th and the $90$th percentile in each distribution.
\textbf{b}: the thin lines represent the$\lambda_{ij}$ values of
a random subset of $200$ links in the case with $M=200$ cascades
as a function of iterations of the GABP algorithm; black thick line:
area under the \emph{ROC} curve.}
\end{figure*}

As can be seen in Fig. \ref{fig:lambda_dynamics} (\textbf{a}), the
distribution of inferred values of true links rapidly separates from
the one of non-existent ones, that concentrates around vanishing values
even for a very small number of observations. The strict separation
of the two distributions confirms the results from the area under
the ROC curve.

It is worth noting that GABP achieves a good level of reconstruction
accuracy in a very small number of steps. The dynamics of the inferred
$\lambda_{ij}$ as a function of iterations of the algorithm is exemplified
in Fig. \ref{fig:lambda_dynamics} (\textbf{b}). Even after a very
small number of iterations, true links are clearly distinguished from
non existent ones, as can be seen from the steep rise of the area
under the ROC curve as a function of iterations: we observe that this
kind of behavior is quite general and not restricted to the case $M=\mathcal{O}\left(N\right)$.

\subsection*{Reconstructing real networks}

\begin{figure*}
\begin{centering}
\includegraphics[width=0.9\columnwidth]{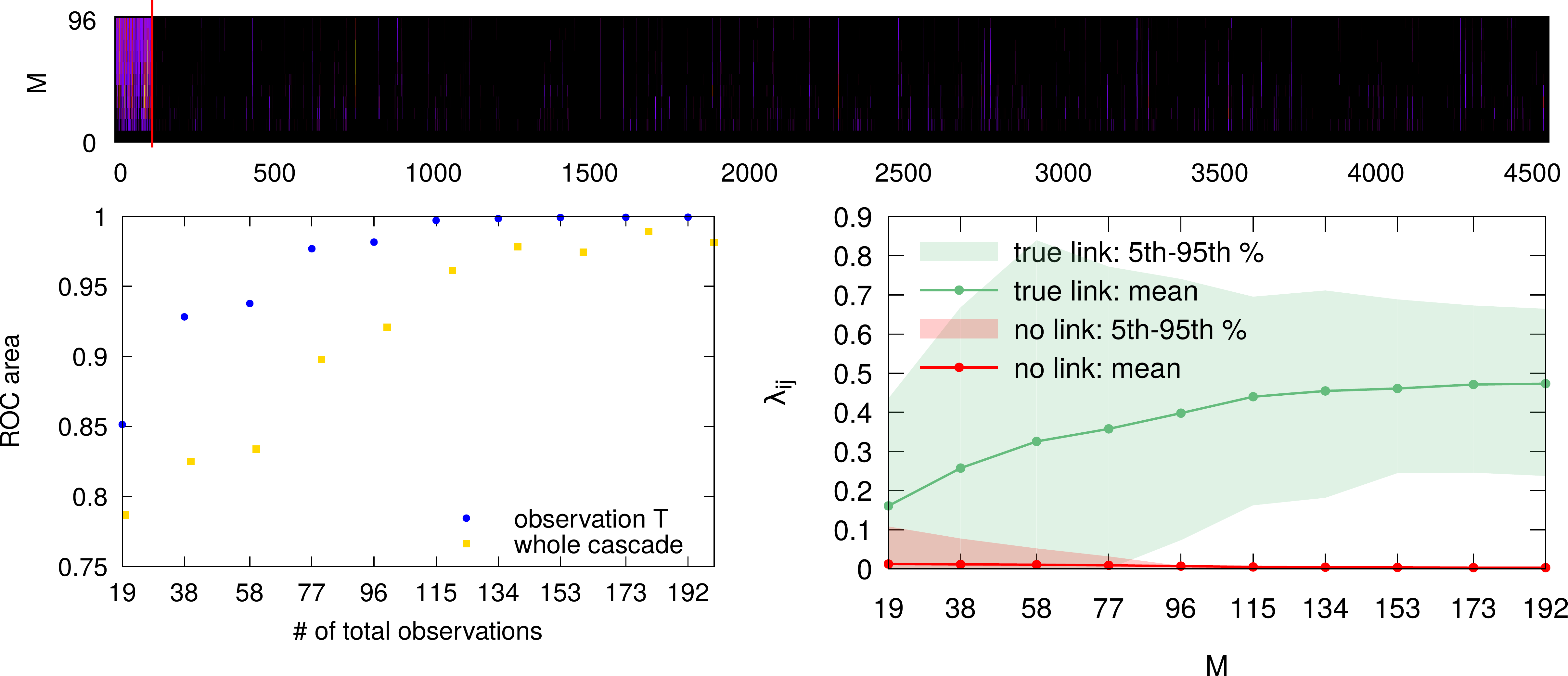}
\par\end{centering}
\centering{}\caption{\label{fig:retweet}Reconstruction performance of GABP in the network
of retweets ($|G|=96$) with increasing number of independent cascades
$M$. Epidemic parameters are $\lambda_{ij}=0.5$ and $\mu_{i}=0.4$,
observation time $T=5$. Gold curve: area under \emph{ROC} curve in
the case where the state of the networks is fully observed at each
time $t\in\left\{ 1,\dots,T\right\} $ for each cascade $m$. Blue
curve: area under ROC curve in the case where the network is observed
only at time $T$ in each cascade. \emph{Inset}: average value of
$\lambda_{ij}$ for true links (green) versus non existent ones (red)
as a function of the number of observed cascades in the standard case
(observation at time $T$ only); shaded areas correspond to the intervals
between the $5$th and the $95$th percentile in each distribution.}
\end{figure*}

We tested the GABP algorithm on two different real interaction networks
on which information about contacts is available for validation purposes.
The first dataset consists of a networks of Twitter retweets \citep{nr:rt-retweet,nr}:
the networks is composed of $|G|=96$ nodes, which represent Twitter
users, linked through $|E|=117$ edges corresponding to retweets (these
were collected from various social and political hash-tags). The average
degree of a node in the network is $d_{av}=2$, with a minimum degree
of $1$ and a maximum degree of $17$. Figure \ref{fig:retweet} shows
the reconstruction performance in the retweet networks using two different
observation paradigms: in the single-observation-per-cascade paradigm
(which we considered as the standard case), the nodes state is available
only once per cascade, whereas in the whole-cascade paradigm all nodes
are observable at all times. It is apparent that an extremely accurate
reconstruction is achievable with a number of cascade $M$ quite small
compared to $|G|$.

\begin{figure}
\begin{centering}
\begin{tabular}{cccc}
\includegraphics[width=0.2\columnwidth]{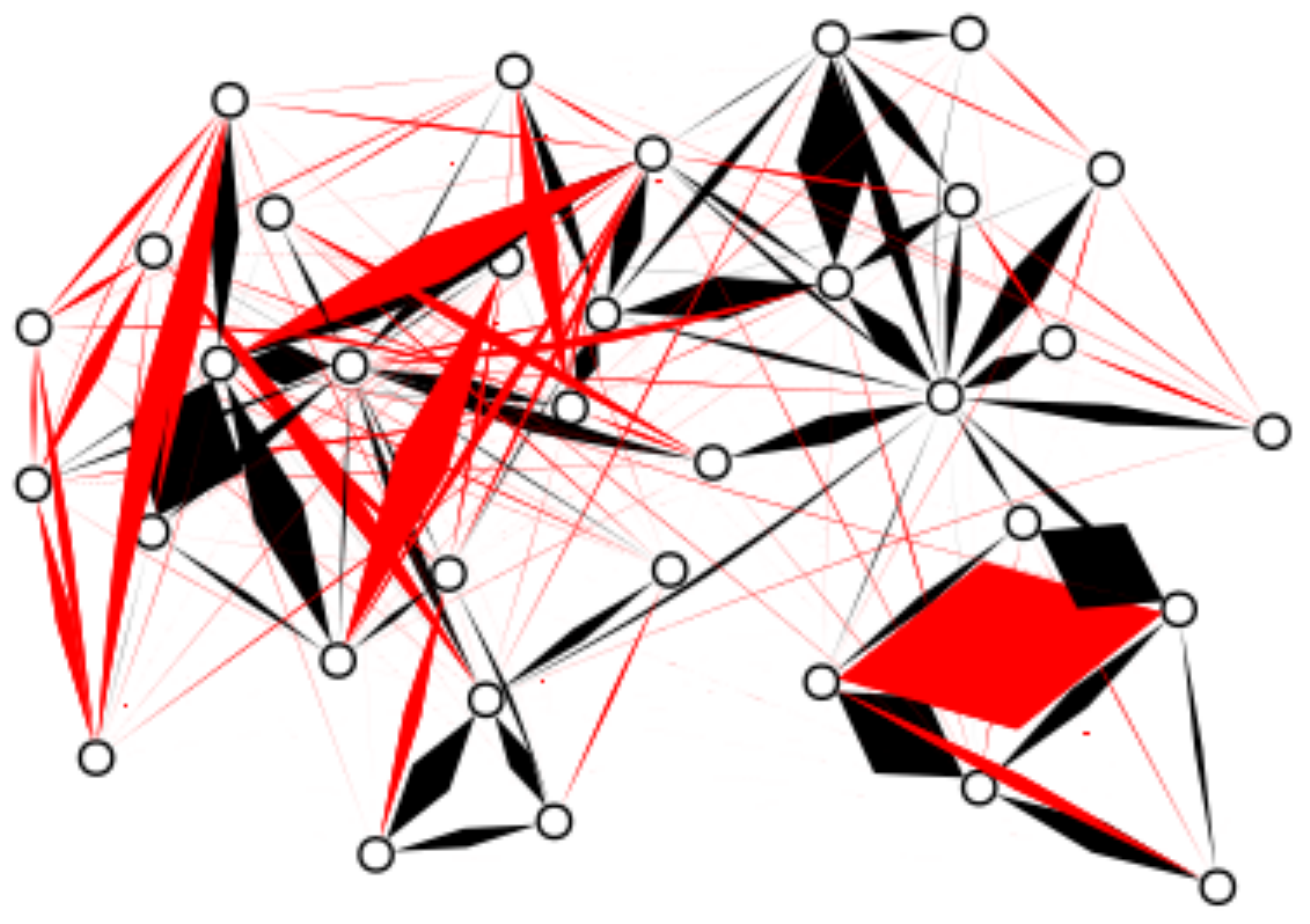} & \includegraphics[width=0.2\columnwidth]{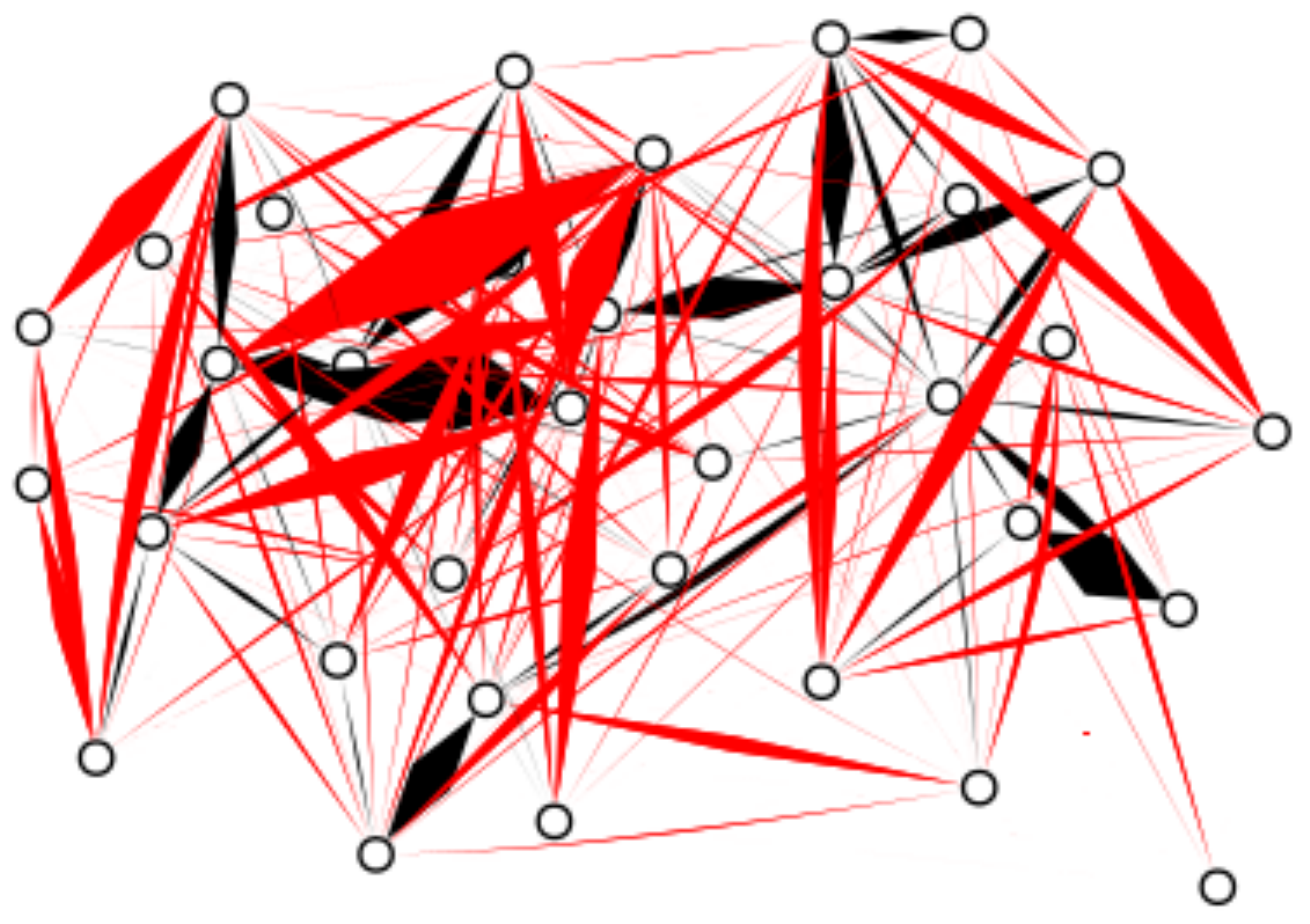} & \includegraphics[width=0.2\columnwidth]{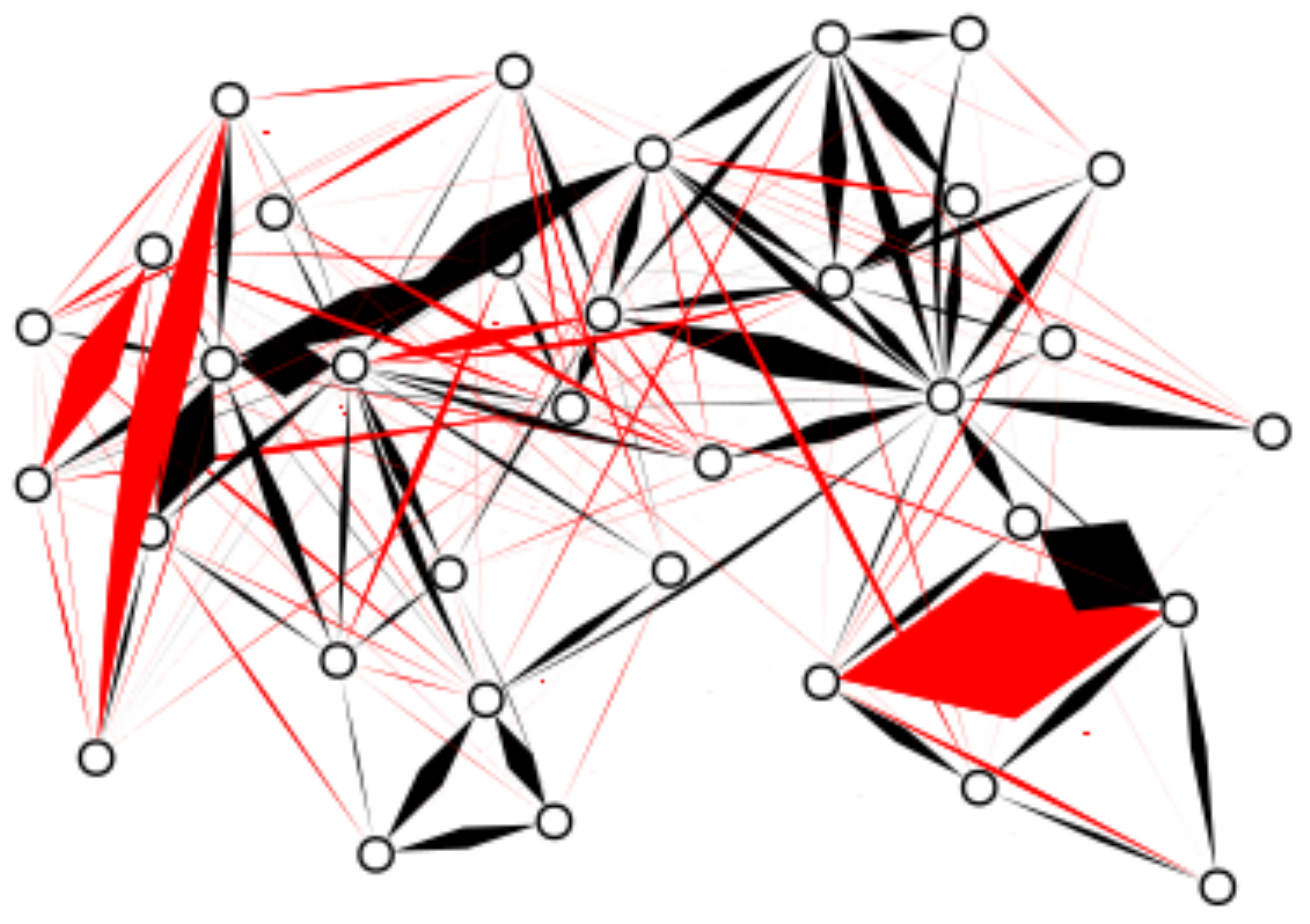} & \includegraphics[width=0.2\columnwidth]{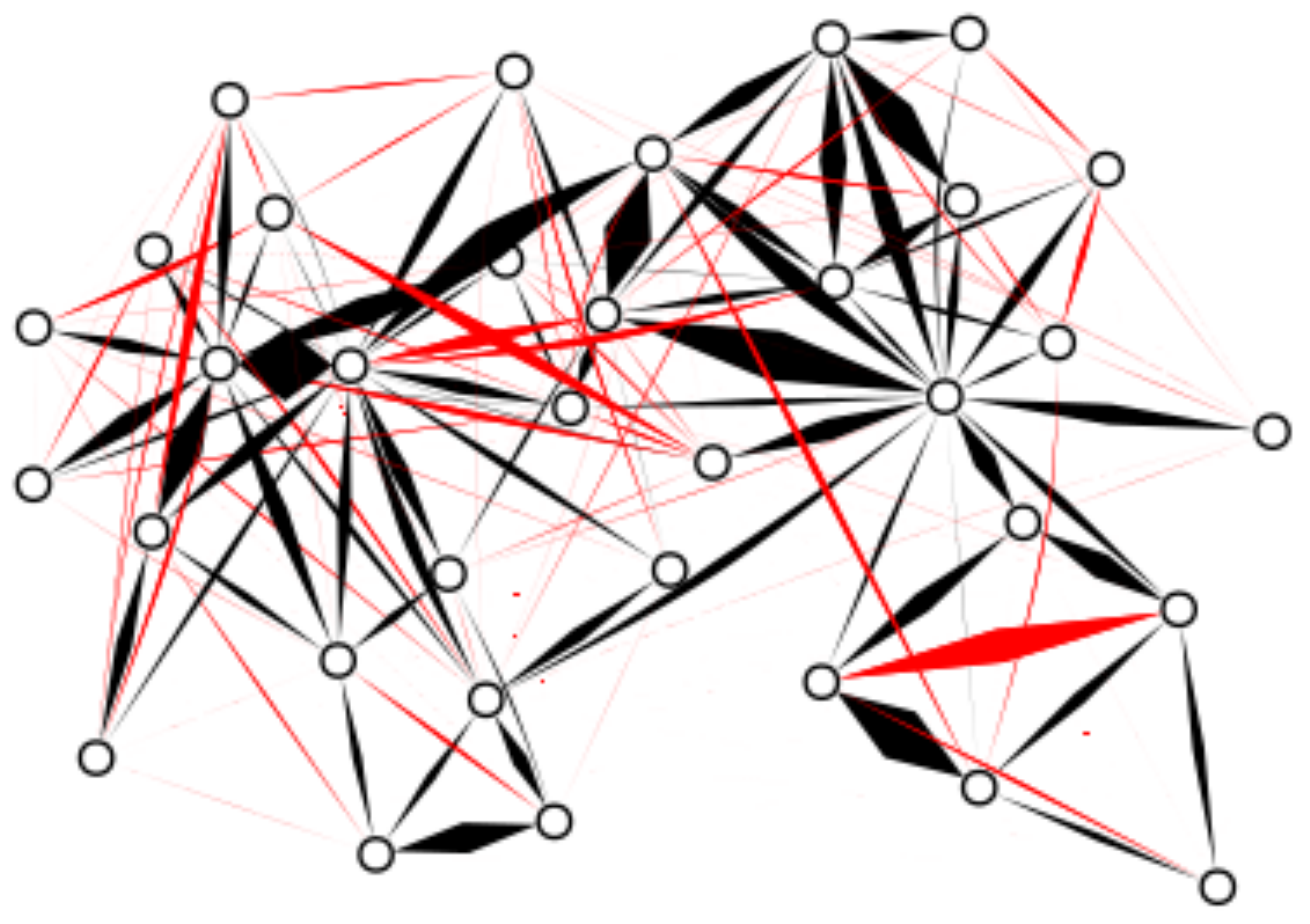}\tabularnewline
$M=14$ & $M=41$ & M=68 & $M=102$\tabularnewline
\end{tabular}
\par\end{centering}
\caption{\label{fig:karate_res_pictorial}Pictorial representation of the GABP
performance in Zachary's Karate Club network with an increasing number
of cascades $M$. An edge is thrown between node $i$ and node $j$
if $\lambda_{ij}$ in non zero, the width of the edge being proportional
to the value $\lambda_{ij}$. True links are colored in black, red
links are not present in the original network.}
\end{figure}

As another illustrative example, in Fig. \ref{fig:karate_res_pictorial}
we show a pictorial representation of the reconstruction of the Zachary's
Karate Club network, a small social network which consists of $|G|=34$
nodes and $|E|=78$ edges, documenting the pairwise interactions over
the course of three years among members of an university-based karate
club. In this case, we simulated up to $M=102$ cascades and investigated
the performance of the inference method with homogeneous parameters
$\lambda=0.3$ and $\mu=0.4$ at increasing $M$. In Fig. \ref{fig:karate_res_pictorial},
links not present in the actual graph are colored in red, and appear
clearly distinguished from the true ones (colored in black) even for
very small values of $M$.

\begin{figure}
\centering{}\includegraphics[width=0.9\columnwidth]{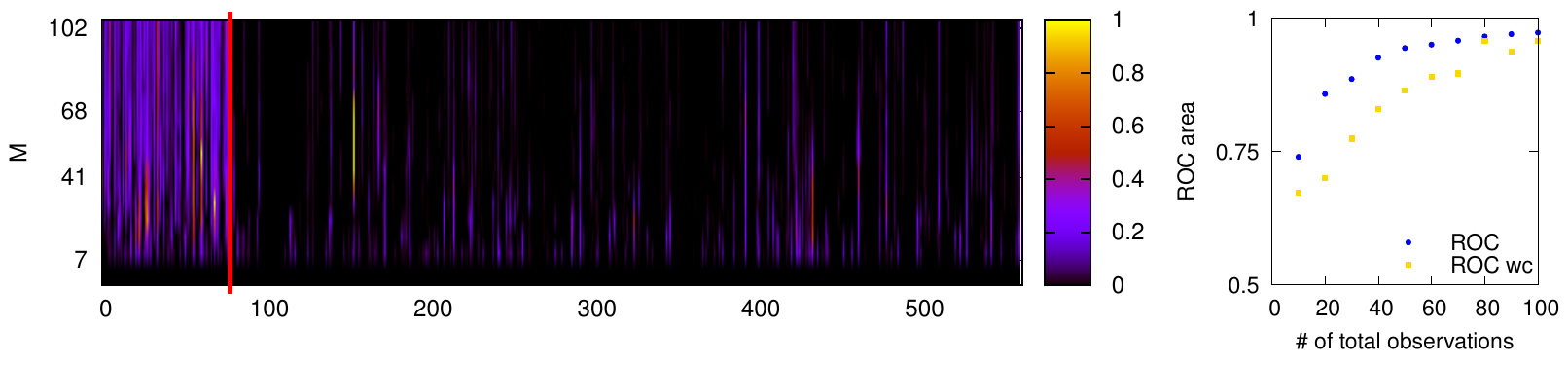}\caption{\label{fig:karate_res}Left: Reconstruction performance of GABP in
the Zachary's Karate Club network with different numbers $M$ of independent
cascades. $M$ is on the $y$ axis. The links are on the $x$ axis,
ordered in such a way that the first $78$ are the true links in the
original graph. The color intensity is proportional to the value $\lambda_{ij}$
for each putative link $\left(i,j\right)$ at increasing values of
$M$. Right: area under the \emph{ROC} curve (\emph{x} axis) for increasing
total observations (see text) of the entire networks (\emph{y} axis,
scale as in the left part). The blue curve corresponds to a single
final observations per cascade at time $T=5$, the gold curve shows
the case in which cascades are fully observed.}
\end{figure}

For a more thorough representation of the reconstruction process in
the Karate Club network, we show in Fig. \ref{fig:karate_res} (Left)
a color intensity plot of the dynamics of inference as the number
of cascades is increased: true links are immediately identified, as
the ROC area indicates (Right, blue curve).

It is very interesting to note that, while observing cascades in their
entirety clearly conveys a lot of information on the network structure,
if the total number of observations of the full state of the network
is constrained, distributing these observations far apart in time
pays better. This is clearly shown in Fig. \ref{fig:karate_res} (Right)
by the difference in the area under the ROC curve between the whole
cascade scenario and the single-observation-per-cascade paradigm.

\subsection*{Detecting false positive links in PPI networks}

A challenging problem in reconstructing protein-protein interaction
networks consists in discriminating between true positive (TP) and
false positive (FP) links. We show in this section how GABP algorithm
can be used as a post-processing method to tackle this issue.

\begin{figure}
\begin{centering}
\includegraphics[width=1\textwidth]{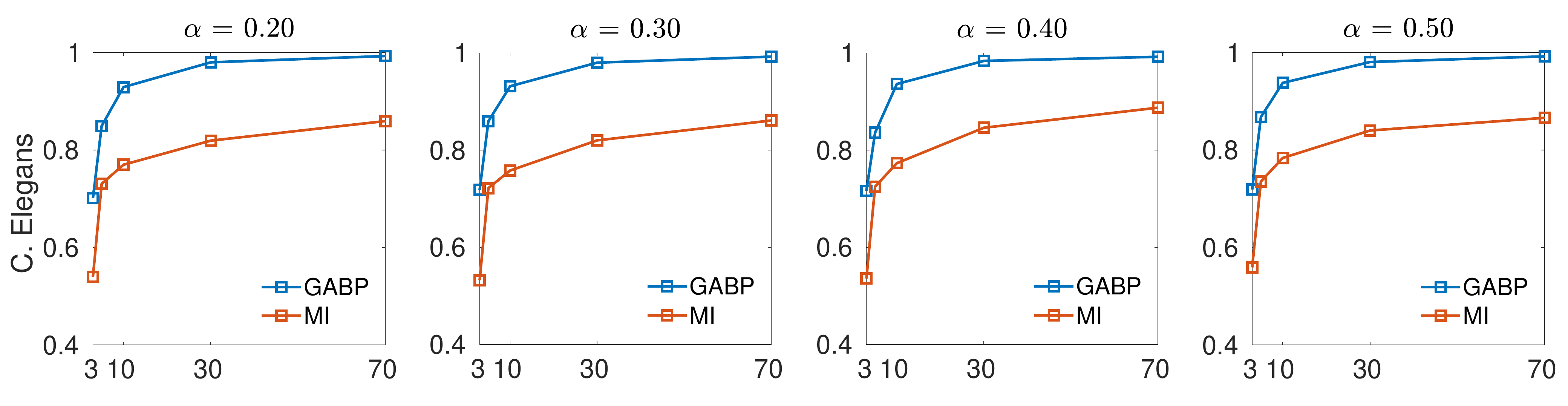}
\par\end{centering}
\begin{centering}
\includegraphics[width=1\textwidth]{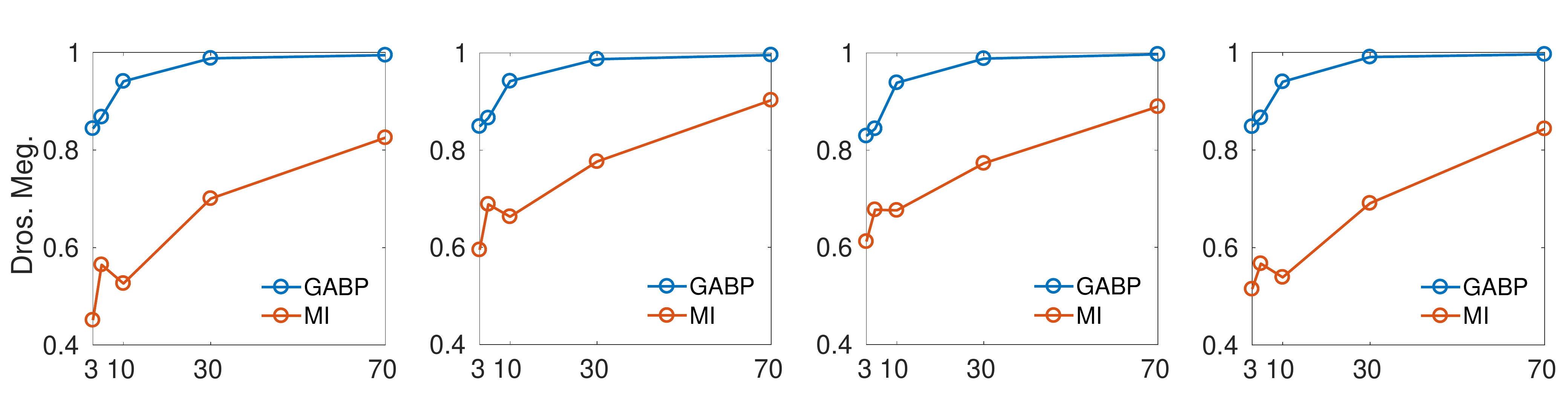}
\par\end{centering}
\begin{centering}
\includegraphics[width=1\textwidth]{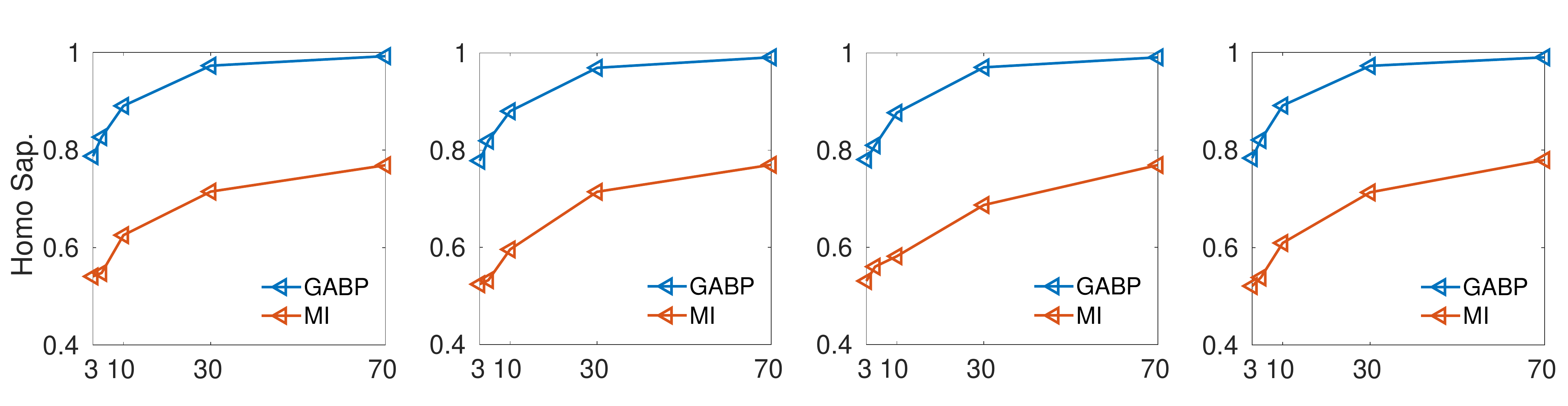}
\par\end{centering}
\begin{centering}
\includegraphics[width=1\textwidth]{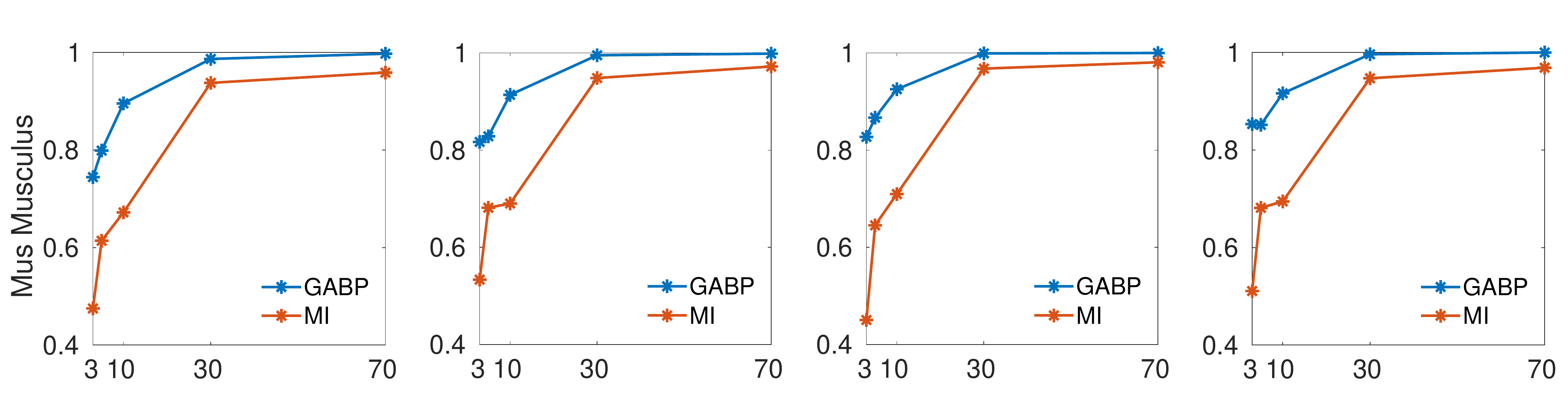}
\par\end{centering}
\begin{centering}
\includegraphics[width=1\textwidth]{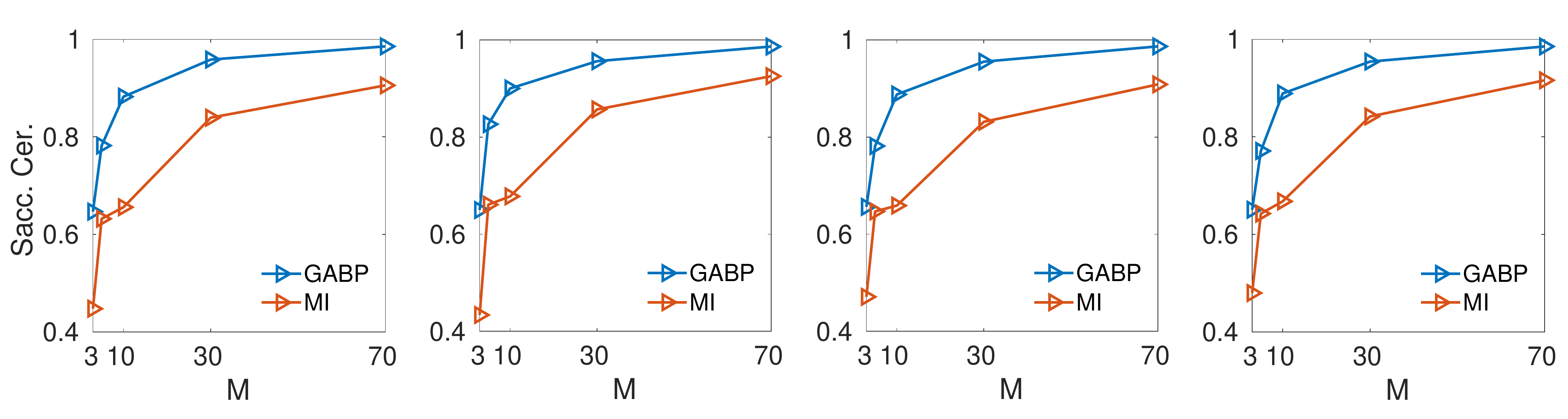}
\par\end{centering}
\caption{\label{fig:ROCareasPPI}Plots of the \textit{ROC} areas for GABP and
MI interactomes predictions. Each row of the table corresponds to
one of the five studied interactomes while each column to a different
$\alpha$, the fraction of extra edges. Subplots show the areas under
the \textit{ROC} curves as a function of the number of cascades $M$
for GABP (blue line) and MI (red line).}
\end{figure}

In our experiments we consider as \textit{ground-truth} networks the
giant components of five interactomes of the PSICQUIC dataset \citep{orchard_molecular_2012}
available on the software Cytoscape 3.5.1 \citep{Shannon2003} (properties
are summarized in Table (\ref{tab:properties-ppi})), while contact
cascades are synthetically simulated with infection parameters $\lambda=0.8$
and $\mu=0.3$. To the true networks we add $Z=\alpha\left|E\right|$
extra edges, for $\alpha=\left[0.2,0.5\right]$, that mimic the presence
of false positive interactions. This step is performed in a ``scale-free''
fashion: we first pick a node $i$ with probability proportional to
its degree and we then connect it to a randomly uniformly chosen node
$j\notin\partial i$. We then simulate $M\in\left[3,\,150\right]$
cascades on the true network and, from the final observations (at
time $T=5$), we try to infer the transmission parameters $\lambda_{ij}$
associated with both true and false positive edges of the extended
graph that, differently to the cases examined before, is not a fully
connected graph. We compare our reconstructions to the ones obtained
by a MI based method. In Fig. \ref{fig:ROCareasPPI} we plot a table
containing the areas under the \textit{ROC} curves as a function of
the number of cascades of the five interaction networks. Each row
of the main figure corresponds to an organism and the columns run
over $\alpha$. For all organisms the areas under the \textit{ROC}
curves of GABP results are significantly larger then those of MI reconstructions
and they reach values above $0.9$ even when few cascades are available,
i.e. $M=10$. Quite surprisingly, performances seem to be independent
on the number of extra-edges suggesting that our method is quite robust
in detecting false positive links when the extended graph to be pruned
has a reasonable, but large, number of edges.

\begin{figure}
\begin{centering}
\includegraphics[width=1\textwidth]{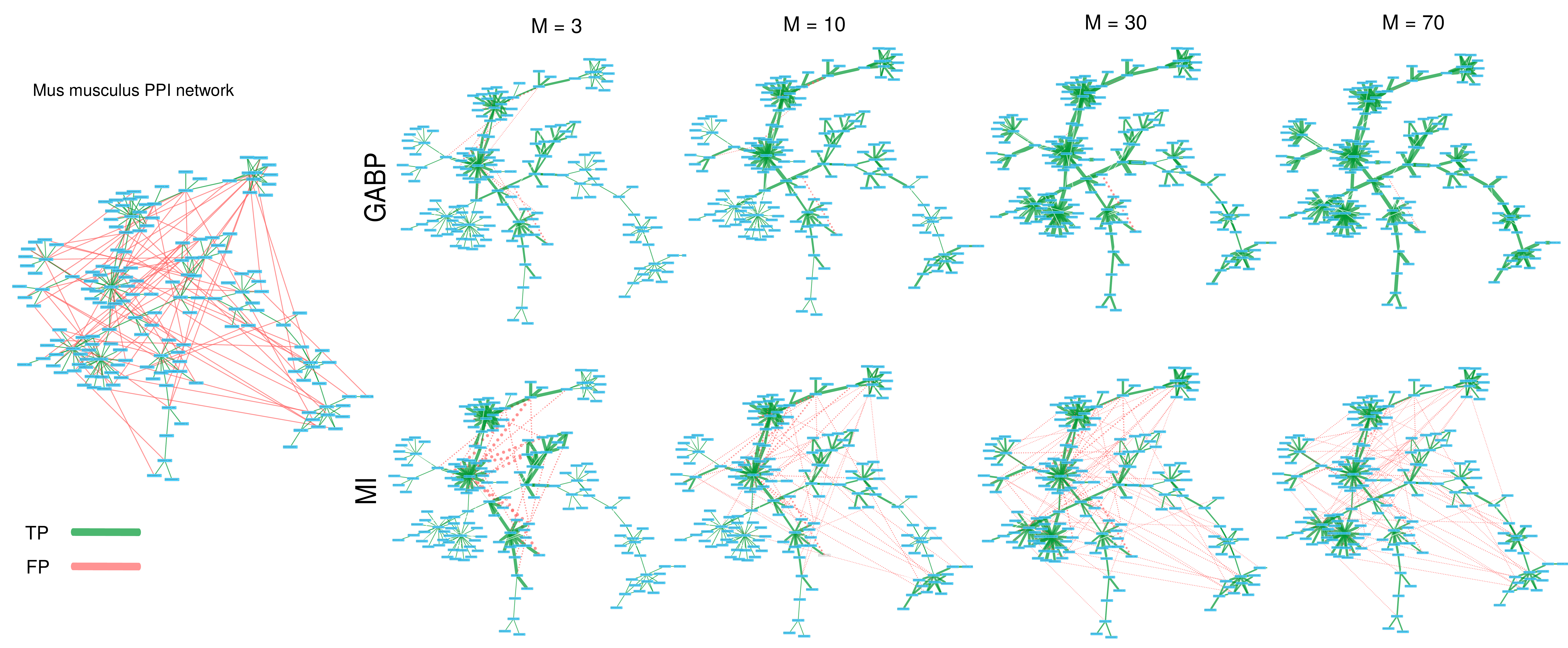}
\par\end{centering}
\caption{\label{fig:False-positive-detection}False positive edges detection
in PPI networks. (\textbf{a}) Mouse interactome of 172 nodes and 297
edges (217 true positive in green and 80 false positive in red). (\textbf{b})
the first (second) row shows the networks reconstructed by GABP (MI)
for $M=\left\{ 3,\,10,\,30,\,70\right\} $. The thickness of each
edge is proportional to the infection parameters of GABP and the mutual
information among couples of nodes for MI; edges with weights smaller
then $10^{-3}$ are not shown.}
\end{figure}

To underline the performances of GABP, we show in Fig. \ref{fig:False-positive-detection}
(\textbf{a}) the \textit{Mus musculus} interactome containing the
true positive (green links) and 80 false positive edges (red links).
The retrieved network for an increasing number of cascades is plotted
in Fig. \ref{fig:False-positive-detection} (\textbf{b}); edges thickness
is proportional to the inferred values of $\lambda_{ij}$ for GABP
and to $m_{ij}$ for MI. It is worth noting that, for very few cascades
($M=3$), both GABP and MI are able to recognize almost all true links
but GABP misclassifies fewer false positive than MI. When $M$ increases,
GABP detects all true edges as the associated $\lambda_{ij}$ significantly
increase and it incorrectly classifies only few false positive edges
that, in any case, exhibit values of the infection parameters close
to zero and negligible if compared to the ones associated with TP
links. On the contrary MI distributes the weights over all the edges
and, for large $M$, it is not able to sharply distinguish the two
sets of links as some of the FP edges have comparable values of $m_{ij}$
to those of TP links.

\begin{table}
\begin{centering}
\begin{tabular}{c|c|c|c}
Organism & |V| & |E| & Dataset name\tabularnewline
\hline 
\hline 
Caenorhabditis Elegans & 372 & 400 & MINT \citep{orchard_mintact_2014}\tabularnewline
\hline 
Drosophila Melanogaster & 398 & 491 & MINT \tabularnewline
\hline 
Homo Sapiens & 801 & 1190 & BHF-UCL\tabularnewline
\hline 
Mus Musculus & 172 & 217 & EBI-GOA-miRNA\tabularnewline
\hline 
Saccharomyces Cerevisiae & 185 & 1476 & UniProt \citep{the_uniprot_consortium_uniprot:_2017}\tabularnewline
\end{tabular}
\par\end{centering}
\caption{\label{tab:properties-ppi}Properties of the interactomes. This table
shows the name of the organisms, the number of nodes and edges of
the PPI networks and the name of the public datasets supported by
PSICQUIC.}
\end{table}

\subsection*{Inferring transmission probabilities}

Let us now briefly consider a slightly different application of the
general formalism presented so far. Suppose that the underlying network
structure is known but little or any information is available on the
transmission probabilities $\lambda_{ij}$, which are, in the general
case, inhomogeneous. Our method can be easily accommodated so as to
provide the maximum likelihood estimation of the quantities $\lambda_{ij}$.
Starting from an initial assignment of the coupling parameters (we
used $\lambda_{ij}\equiv0.5)$ defined over a known topology, one
seeks a fixed point of the coupled BP and gradient equations using
GABP.

\begin{figure}
\centering{}\includegraphics[width=1\columnwidth]{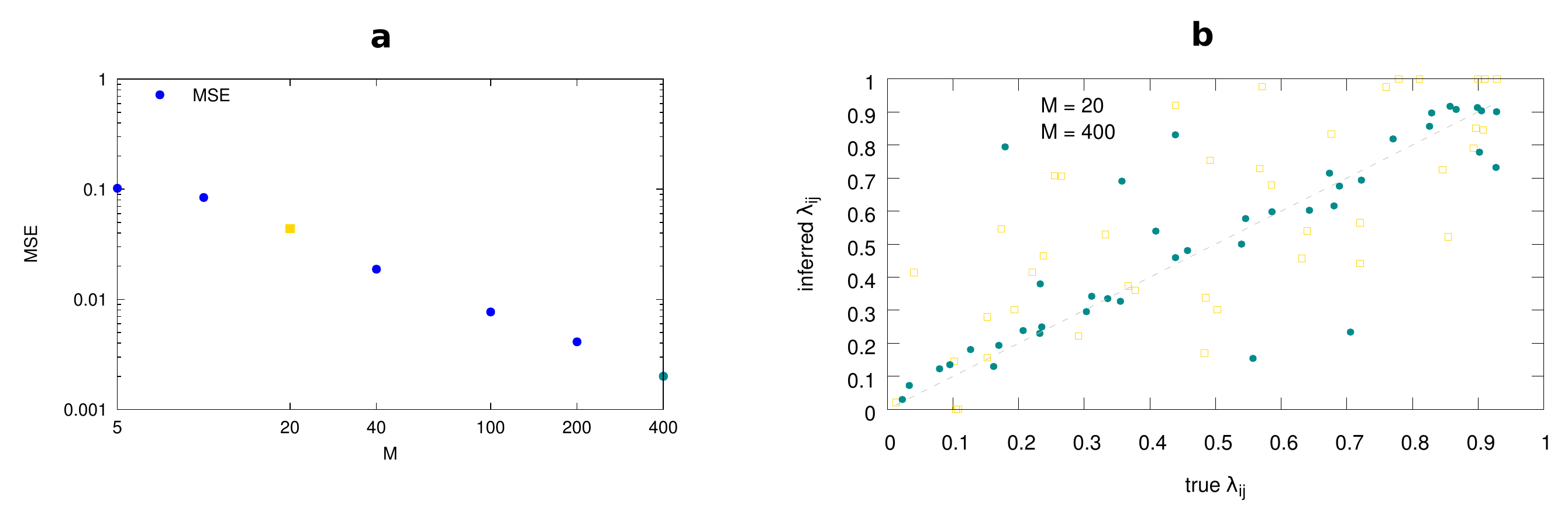}\caption{\label{fig:spreading_coupling}Reconstructing spreading couplings
in inhomogeneous networks. \textbf{a}: mean squared reconstruction
error $MSE=\frac{\sum_{i<j}\left(\lambda_{ij}-\lambda_{ij}^{true}\right)^{2}}{|E|}$
in a random regular graph of size $|G|=20$ and degree $d=4$, as
a function of the number of observed cascades $M$. The network structure
is known in advance. The spreading couplings $\lambda_{ij}^{true}$
have been extracted randomly from the homogeneous distribution in
the interval $[0,1]$. The state of the network is observed only at
time $T=5$ for each cascade. \textbf{b}: scatter plot of reconstructed
transmission probabilities $\lambda_{ij}$ versus true spreading couplings
$\lambda_{ij}^{true}$ for the cases $M=20$ and $M=400$, corresponding
to the golden and green points in the left plot, respectively.}
\end{figure}

As an example, we consider a random regular graph of size $|G|=20$
with degree $d=4$, and evaluate the inference performance with increasing
number of cascades $M$. Figure \ref{fig:spreading_coupling} (\textbf{a})
shows the value of the Mean Square Error $MSE=\frac{\sum_{\left(ij\right)\in E}\left(\lambda_{ij}-\lambda_{ij}^{true}\right)^{2}}{|E|}$
between the inferred transmission probabilities $\lambda_{ij}$ and
the true ones, $\lambda_{ij}^{true}$. To better appreciate the quality
of the inference, we show a scatter plot for two different values
of $M$ in Figure \ref{fig:spreading_coupling} (\textbf{b}).

\section*{Discussion}

We have presented a new method that allows to reconstruct a hidden
network from limited information of activity propagations, and showed
that the reconstruction performance is extremely accurate even when
the number of snapshot observations is very small. This scheme can
be effectively applied to the detection of false positive links in
protein-protein interactions networks even when the number of candidate
false edges is comparable to the effective number of true positive
contacts. In this particular case it suffices very few independent
cascades to correctly classify the great majority of the links.

There are several advantages of this approach over existing ones.
The main one is that several inference problems can be treated under
a unique formulation. Our technique can be easily extended to incorporate
effects of unreliable observations, taking into account all those
situations when some noise enters the measurements, or all those cases
where Susceptible nodes cannot be distinguished from Recovered ones
\citep{altarelli_patient-zero_2014}. When a complete list of contact
times between nodes is available, the construction of an equivalent
network of timely dependent infection probability is straightforward,
and the current approach has been proven to be effective. 

Owing to the generality of the Bayesian method, the described technique
is capable of dealing with a wide variety of irreversible spreading
processes on networks. A possible simple generalization is to the
(random) Bootstrap Percolation case where each node gets activated
when aggregated input from neighbors overcome an intrinsic stochastic
activation threshold of the node. These models are widely used to
describe the features of dynamical processes in neuronal networks,
and we consider this an exciting research direction.

\section*{Methods}

\subsection*{Graphical model formulation of the spreading process}

Let us first consider a single cascade on a network with a fixed topology.
For a fixed initial configuration $\mathbf{x}\left(0\right)$, a realization
of the stochastic process can be generated by drawing randomly a set
of infection transmission delay $s_{ij}$ for all pairs $(ij)$ and
the recovery times $g_{i}$ of each node $i$. The recovery times
$\{g_{i}\}$ are independent random variables extracted from the geometric
distributions $\mathcal{G}_{i}\left(g_{i}\right)=\mu_{i}\left(1-\mu_{i}\right)^{g_{i}}$,
the delays $\{s_{ij}\}$ are conditionally independent random variables
distributed according to a truncated geometric distribution, 
\begin{equation}
\omega_{ij}\left(s_{ij}|g_{i}\right)=\begin{cases}
\lambda_{ij}\left(1-\lambda_{ij}\right)^{s_{ij}}, & s_{ij}\leq g_{i}\\
\left(1-\lambda_{ij}\right)^{g_{i}+1}, & s_{ij}=\infty,
\end{cases}\label{eq:decay}
\end{equation}
Note that we concentrate in the value $s_{ij}=\infty$ the mass of
the distribution beyond the hard cut-off $g_{i}$ imposed by the recovery
time. The joint probability distribution of infection and recovery
times conditioned on the initial state is easily written: 
\begin{align}
\mathcal{P}\left(\mathbf{t},\mathbf{g}|\mathbf{x}\left(0\right)\right) & =\sum_{\mathbf{s}}\mathcal{P}\left(\mathbf{s}|\mathbf{g}\right)\mathcal{P}\left(\mathbf{t}|\mathbf{x}\left(0\right),\mathbf{s},\mathbf{g}\right)\mathcal{P\left(\mathbf{g}\right)}\nonumber \\
 & =\sum_{\mathbf{s}}\prod_{i,j}\omega_{ij}\left(s_{ij}|g_{i}\right)\prod_{i}\psi_{i}(t_{i},\{t_{k},s_{ki}\}_{k\in\partial i})\mathcal{G}_{i}(g_{i}),\label{eq:direct}
\end{align}
where 
\begin{equation}
\psi_{i}(t_{i},\{t_{k},s_{ki}\}_{k\in\partial i})=\delta(t_{i,}\mathbb{I}[x_{i}\left(0\right)\neq I](1+\min_{k\in\partial i}\{t_{k}+s_{ki}\}))
\end{equation}
is a characteristic function which imposes on each node $i$ the dynamical
constraint of equation (\ref{eq:dynamical}).

Using the Bayes formula, the posterior probability of the initial
configuration given an observation at time $T$ reads:

\begin{eqnarray}
\mathcal{P}\left(\mathbf{x}\left(0\right)|\mathbf{x}\left(T\right)\right) & \propto & \sum_{\mathbf{t,g}}\mathcal{P}\left(\mathbf{x}\left(T\right)|\mathbf{t},\mathbf{g}\right)\mathcal{P}\left(\mathbf{t},\mathbf{g}|\mathbf{x}\left(0\right)\right)\mathcal{P}\left(\mathbf{x}\left(0\right)\right)\\
 & = & \sum_{\mathbf{t},\mathbf{g},\mathbf{s}}\prod_{i,j}\omega_{ij}\prod_{i}\psi_{i}\mathcal{G}_{i}\gamma_{i}\zeta_{i}^{T}\label{eq:bayes}
\end{eqnarray}
where $\mathcal{P}\left(\mathbf{x}\left(0\right)\right)=\prod_{i}\gamma_{i}\left(x_{i}\left(0\right)\right)$
is a factorized prior on the initial infection with 
\begin{equation}
\gamma_{i}(x_{i}\left(0\right))=\gamma\delta(x_{i}\left(0\right),I)+(1-\gamma)\delta(x_{i}\left(0\right),S)\label{eq:prior}
\end{equation}
for a generally small constant $\gamma$ (we don't allow state $(R)$
at time 0). Note that the network state $\mathbf{x}\left(t\right)$
is a deterministic function of the set of infection and recovery times
$(\mathbf{t},\mathbf{g})$, so that we obtain

\begin{eqnarray}
\mathcal{P}\left(\mathbf{x}\left(T\right)|\mathbf{t},\mathbf{g}\right) & = & \prod_{i}\zeta_{i}^{T}\left(t_{i},g_{i},x_{i}\left(T\right)\right)
\end{eqnarray}
with $\zeta_{i}^{t}=\mathbb{I}\left[x_{i}\left(t\right)=S,t<t_{i}\right]+\mathbb{I}\left[x_{i}\left(t\right)=I,t_{i}\leq t<t_{i}+g_{i}\right]+\mathbb{I}\left[x_{i}\left(t\right)=R,t_{i}+g_{i}\leq t\right].$
Note that assuming $x_{i}\left(0\right)\in\left\{ \left(S\right),\left(I\right)\right\} $,
then $\psi_{i}(t_{i},\{t_{k},s_{ki}\}_{k\in\partial i})$ could be
also rewritten equivalently as $\zeta_{i}^{0}\left(t_{i},g_{i},x_{i}\left(0\right)\right)[\delta(t_{i,}1+\min_{k\in\partial i}\{t_{k}+s_{ki}\})+\delta\left(t_{i},0\right)]$.
Now, if we introduce a set of observational weights $\zeta_{i}^{m,T}$,
one for each observation $m$, together with a set of priors priors
$\zeta_{i}^{m,0}$, the posterior distribution of the initial states
conditioned to observations, because of the assumption of independence,
will be proportional to the product over all the single probability
weights for each cascade $\mathcal{P}\left(\mathbf{x}^{1:M}\left(0\right)|\mathbf{x}^{1:M}\left(T\right)\right)\propto\prod_{m=1}^{M}\sum_{\mathbf{t}^{m},\mathbf{g}^{m}}\mathcal{P}\left(\mathbf{x}^{m}\left(T\right)|\mathbf{t}^{m},\mathbf{g}^{m}\right)\mathcal{P}\left(\mathbf{t}^{m},\mathbf{g}^{m}|\mathbf{x}^{m}\left(0\right)\right)\mathcal{P}\left(\mathbf{x}^{m}\left(0\right)\right)$
that taking into account equation (\ref{eq:bayes}) will take the
form:
\begin{eqnarray}
\mathcal{P}\left(\mathbf{x}^{1:M}\left(0\right)|\mathbf{x}^{1:M}\left(T\right)\right) & \propto & \prod_{m=1}^{M}\sum_{\mathbf{t}^{m},\mathbf{g}^{m},\mathbf{s}^{m}}\prod_{i<j}\omega_{ij}^{m}\prod_{i}\psi_{i}^{m}\mathcal{G}_{i}^{m}\gamma_{i}^{m}\zeta_{i}^{m,T}\label{eq:idnr_graphical_model}
\end{eqnarray}
where all the factors have been labeled with an extra cascade index
$m$ and $\mathbf{x}^{1:M}\left(T\right)=\left(\mathbf{x}^{m}\left(T\right)\right)_{m=1,\dots,M}$.
Since we have no \emph{a priori} information on the graph topology,
the product in the term $\prod_{i<j}\omega_{ij}^{m}$ runs over all
the possible pair $i$ and $j$ in the set $V$, meaning that we always
work in the setting of a fully connected network with weights $\left\{ \lambda_{ij}\right\} $.
If the number of cascades $M$ is large enough, the non zero elements
of the matrix $\left\{ \lambda_{ij}\right\} $ will signal, upon convergence
of the GABP algorithm, the true links in the original graph, their
value being informative of the heterogeneity of infection probabilities.
The same holds for the set of recovery parameters $\left\{ \mu_{i}\right\} $.
Note that for $\lambda_{ij}=0$, (\ref{eq:decay}) imposes the condition
$s_{ij}=\infty$, meaning that $\left(ij\right)$ can be ignored in
(\ref{eq:dynamical}), effectively pruning the link from the equations.

\subsection*{Belief Propagation approach}

Given a high dimensional probability distribution $M\left(\mathbf{z}\right)$
with a locally factorized interaction structure, computing marginals
and aggregated quantities may be addressed with the use of a Message
Passing procedure built on a cavity approximation for locally tree-like
graphs \citep{yedidia_bethe_2001,Yedidia:2003:UBP:779343.779352,mezard_information_2009}.
In the present problem, we obtain a full set of (cavity) marginal
probabilities over the set of all the possible cascades compatible
with the observations. BP is proven to be exact on tree graphs, and
has been successfully employed on general loopy graphs under mild
regularity conditions \citep{altarelli_large_2013,altarelli_optimizing_2013}\citep{altarelli_large_2013,altarelli_optimizing_2013}.

To briefly describe the essence of the the method, let us consider
a probability distribution over the variables $\mathbf{z}=\left\{ z_{i}\right\} $
that has the following factorized form: 
\begin{equation}
M\left(\mathbf{z}\right)=\frac{1}{Z}\prod_{a}\chi_{a}\left(\mathbf{z}_{a}\right)\label{eq:bp_prob_distr}
\end{equation}
where each $\chi_{a}$ is called compatibility function, or \emph{factor}.
We write $\mathbf{z}_{a}=\left\{ z_{i}\right\} _{i\in\partial a}$
as the set of variables it depends on, $\partial a$ the subset of
indices of variables in factor $\chi_{a}$, and accordingly $\partial i$
will be the subset of factors that depend on $z_{i}$. Belief Propagation
equations are a set of self-consistent equations for the so-called
\emph{cavity messages} (or \emph{beliefs}), a set of single-site probability
distributions which are associated to each directed link in the graphical
model representing to the joint distribution of equation (\ref{eq:bp_prob_distr}).
The general form of BP equations is the following: 
\begin{eqnarray}
p_{\chi_{a}\to i}\left(z_{i}\right) & = & \frac{1}{Z_{ai}}\sum_{\left\{ z_{j}:j\in\partial a\setminus i\right\} }\chi_{a}\left(\mathbf{z}_{a}\right)\prod_{j\in\partial a\setminus i}m_{j\to\chi_{a}}\left(z_{j}\right)\label{eq:factor-to-var}\\
m_{i\to\chi_{a}}\left(z_{i}\right) & = & \frac{1}{Z_{ia}}\prod_{b\in\partial i\setminus a}p_{\chi_{b}\to i}\left(z_{i}\right)\label{eq:var-to-factor}\\
m_{i}\left(z_{i}\right) & = & \frac{1}{Z_{i}}\prod_{b\in\partial i}p_{\chi_{b}\to i}\left(z_{i}\right)\label{eq:var-marginal}
\end{eqnarray}
where the terms $Z_{ia},Z_{ai}$ and $Z_{i}$ are local partition
function, serving as normalizers. To solve equations (\ref{eq:factor-to-var})
and (\ref{eq:var-to-factor}) an iterative procedure is typically
used, where the cavity messages are initialized with uniform distributions
and they are asynchronously updated until convergence to a fixed point
(see e.g. \citep{yedidia_bethe_2001,mezard_information_2009} for
an introduction). The BP equations can be thought as local update
rules for messages in a so-called Factor Graph, a bipartite graph
where each term $\chi_{a}$ is associated to a factor node, connected
to all the variable nodes in the set $\mathbf{z}_{a}$ it depends
on. A naive implementation of the BP scheme at the level of equation
(\ref{eq:idnr_graphical_model}) would simply not work, since the
corresponding graphical model has a loopy structure both at local
and global scale. It is however possible to construct a disentangled
factor graph by means of a re-parametrization of the cavity messages.
We provide a brief description of this procedure in the Supplementary
Methods. For a thorough discussion we refer the reader to previous
works (see \citep{altarelli_bayesian_2014}, \citep{altarelli_patient-zero_2014}).
Here we just want to stress that the modified factor graph is an enriched
dual version of the original graph, whence the particular appeal of
the method. In particular, this implies that Belief Propagation provides
the exact Bayesian solution when the underlying network is acyclic.

While the computation of equation (\ref{eq:var-to-factor}) is straightforward,
the sum in equation (\ref{eq:factor-to-var}) generally involves a
number of steps growing exponentially with the size of $\partial a$.
An efficient implementation of the BP equations for the posterior
distribution is given in the Supplementary Methods. Once Belief Propagation
converges, equation (\ref{eq:var-marginal}) can be used to compute
the marginal probability $\mathcal{P}\left(t_{i}^{n}=0\mid\left\{ \mathbf{x}^{m}\left(T\right)\right\} \right)$,
which brings a posterior estimation of the probability for the node
$i$ to be the active at time $t=0$ in the $m$th cascade.

\subsection*{Network reconstruction algorithm}

We employ an alternating optimization scheme in which Belief Belief
Propagation is coupled to a Maximum Likelihood strategy, implemented
with a Gradient Ascent method. In the BP phase, the network parameters
$\left\{ \lambda_{ij},\mu_{i}\right\} $ are kept fixed and a solution
is searched iteratively for equations (\ref{eq:factor-to-var}) and
(\ref{eq:var-to-factor}). At this stage, the source can be located
independently for each cascade looking at the single-site marginals
$\mathcal{P}\left(x_{i}^{m}\left(0\right)|\left\{ \mathbf{x}^{m}\left(T\right)\right\} \right)$.
In the Maximum Likelihood phase, the log-likelihood of network parameters
is maximized by means of a simple Gradient Ascent (GA) procedure.
The gradient may be computed efficiently in the BP approximation.
The likelihood $\mathcal{P}\left(\left\{ \mathbf{x}^{m}\left(T\right)\right\} |\left\{ \lambda_{ij}\right\} ,\left\{ \mu_{i}\right\} \right)$
with respect to the network parameters is
\[
Z\left(\left\{ \lambda_{ij}\right\} ,\left\{ \mu_{i}\right\} \right)=\prod_{m=1}^{M}\sum_{\mathbf{x}^{m}\left(0\right),\mathbf{t}^{m},\mathbf{g}^{m}}\mathcal{P}\left(\mathbf{x}^{m}\left(T\right)|\mathbf{t}^{m},\mathbf{g}^{m}\right)\mathcal{P}\left(\mathbf{t}^{m},\mathbf{g}^{m}|\mathbf{x}^{m}\left(0\right)\right)\mathcal{P}\left(\mathbf{x}^{m}\left(0\right)\right)
\]
The logarithm of this quantity (log-likelihood) corresponds to the
negative free energy of the model $\mathcal{L}\left(\left\{ \lambda_{ij}\right\} ,\left\{ \mu_{i}\right\} \right)=-f\left(\left\{ \lambda_{ij}\right\} ,\left\{ \mu_{i}\right\} \right)=\log Z\left(\left\{ \lambda_{ij}\right\} ,\left\{ \mu_{i}\right\} \right)$,
and can be expressed as a sum of local terms depending only on the
BP messages (see Supplementary Methods). BP updates for the distribution
in equation (\ref{eq:idnr_graphical_model}) are then coupled to Gradient
Ascent (GA) updates with respect to each network parameter, that take
the form: 
\begin{eqnarray}
\lambda_{ij} & \leftarrow & \lambda_{ij}+\epsilon\frac{\partial\mathcal{L}}{\partial\lambda_{ij}}\\
\mu_{i} & \leftarrow & \mu_{i}+\epsilon\frac{\partial\mathcal{L}}{\partial\mu_{i}}
\end{eqnarray}
with $\epsilon$ a small multiplier parameter (we found $\epsilon=10^{-4}$
yields good results and stable convergence and used this value for
all our simulations). The results presented in this work have been
obtained by interleaving one BP step with a GA step: this simple scheme
suffices to provide good joint estimates for the patient zero in each
cascade, together with a remarkably good reconstruction of the underlying
network. An alternative would consist in applying an expectation maximization
(EM) scheme, in which alternatively BP equations are iterated to convergence
(BP step) and parameters are fully optimized for fixed BP messages
(EM step). However, the EM step requires the maximization of a high
order polynomial that must be solved numerically in any case (e.g.
in a GA scheme). We obtained faster convergence by alternating single
GA and BP steps rather than alternating full convergence cycles of
both steps.

\subsection*{Mutual information \label{subsec:Mutual-information}}

For comparison we have tried to reconstruct the networks in interest
using correlation based measures. At the observation time, we have
computed the probabilities of observing edges $\left(i,j\right)$
as the mutual information between nodes $i$ and $j$:
\begin{equation}
m_{ij}=\sum_{\{x_{i},x_{j}\}}f_{ij}\left(x_{i}\left(T\right),x_{j}\left(T\right)\right)\log\frac{f_{ij}\left(x_{i}\left(T\right),x_{j}\left(T\right)\right)}{f_{i}\left(x_{i}\left(T\right)\right)f_{j}\left(x_{j}\left(T\right)\right)}
\end{equation}
where $f_{ij},\,f_{i}\,$ are empirical probabilities computed as
\begin{align}
f_{ij}\left(x_{i}\left(T\right),x_{j}\left(T\right)\right) & =\frac{1}{M}\sum_{m}\delta_{x_{i}\left(T\right),\,x_{i}^{m}\left(T\right)}\delta_{x_{j}\left(T\right),\,x_{j}^{m}\left(T\right)}\\
f_{i}\left(x_{i}\left(T\right)\right) & =\frac{1}{M}\sum_{m}\delta_{x_{i}\left(T\right),\,x_{i}^{m}\left(T\right)}
\end{align}

\global\long\def\dd{\partial}
 \global\long\def\I{\mathbb{I}}
 \global\long\def\sign{\mbox{sign}}

\appendix

\section{BP equation: efficient disentangled implementation}

We would like to use a factor graph representation that maintains
the same topological properties of the original graph of contacts,
in order to guarantee that BP is exact when the original contact graph
is a tree. Following an approach developed in previous works \citep{altarelli_large_2013,altarelli_optimizing_2013,altarelli_patient-zero_2014},
we proceed to disentangle the factor graph by grouping pairs of infection
times $(t_{i},t_{j})$ in the same variable node. For convenience,
we will keep all variable nodes $\{t_{i}\}$ but we will also introduce
for each edge $(i,j)$ emerging from a node $i$ a set of copies $t_{i}^{(j)}$
of the infection time $t_{i}$, that will be forced to take the common
value $t_{i}$ by including the constraint $\prod_{k\in\partial i}\delta(t_{i}^{(k)},t_{i})$
in an additional factor $\phi_{i}$.

The factors $\phi_{i}$ depend on infection times and transmission
delays just through the sums $t_{i}^{(j)}+s_{ij}$, so that it is
more convenient to introduce the variables $t_{ij}=t_{i}^{(j)}+s_{ij}$
and express the dependencies through the pairs $(t_{i}^{(j)},t_{ij})$.

\begin{figure}
\centering{}\includegraphics[width=0.6\columnwidth]{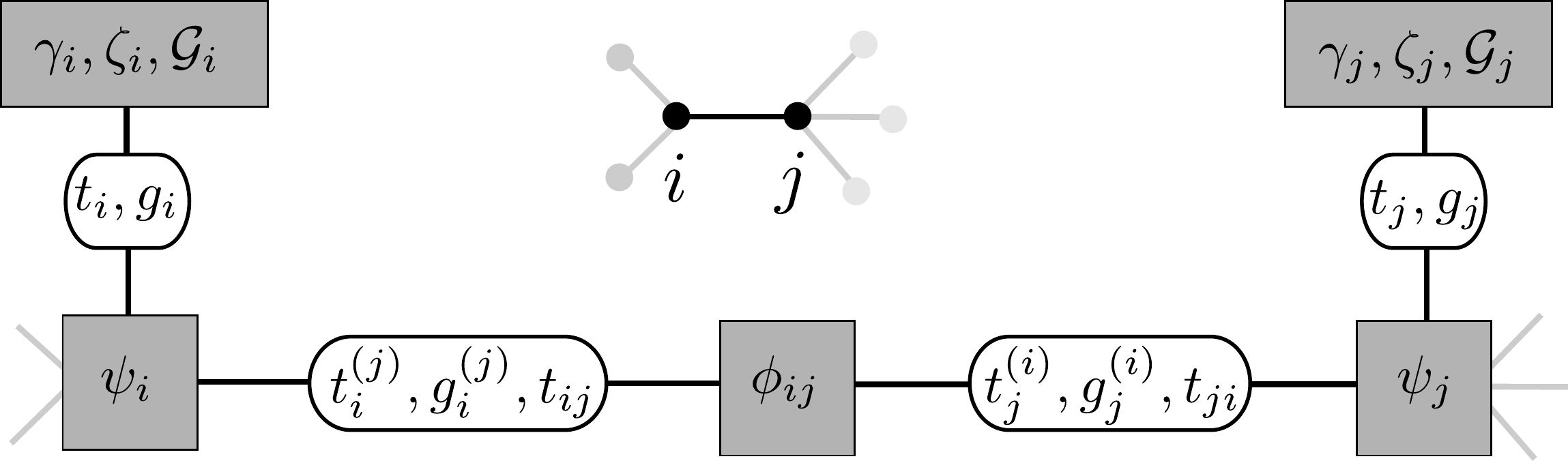}\caption{\label{fig:factor_graph}Disentangled Factor graph representation
of the graphical model. White round nodes correspond to variables,
gray rectangle nodes correspond to factors (or constraints). The topology
of the disentangled factor graph follows the one of the original contact
network.}
\end{figure}

Finally it is convenient to group the variable $g_{i}$ with the corresponding
infection times $t_{i}$ in the same variable node, replace $g_{i}$
and $g_{j}$ by their copies $g_{i}^{(j)}$ and $g_{j}^{(i)}$ in
the edge constraints $\omega_{ij}(t_{ij}-t_{i}^{(j)}|g_{i}^{(i)})$
and $\omega_{ji}(t_{ji}-t_{j}|g_{j}^{(i)})$ and impose the identity
$\prod_{k\in\partial i}\delta(g_{i}^{(k)},g_{i})$ for each node $i$.
The resulting disentangled factor graph appears in Fig. \ref{fig:factor_graph}.

An efficient form for the update equations of the~$\psi_{i}$ factor
nodes is the following: 
\begin{eqnarray}
p_{\psi_{i}\to j}\left(t_{i}^{\left(j\right)},t_{ji},g_{i}^{\left(j\right)}\right) & \propto & \sum_{g_{i},t_{i}}\sum_{\left\{ t_{i}^{\left(k\right)},t_{ki},g_{i}^{\left(k\right)}\right\} }m_{i\to\psi_{i}}\left(t_{i},g_{i}\right)\times\\
 &  & \times\prod_{k\in\partial i\setminus j}m_{k\to\psi_{i}}\left(t_{i}^{\left(k\right)},t_{ki},g_{i}^{\left(k\right)}\right)\psi_{i}\left(t_{i},g_{i},\left\{ \left(t_{i}^{(k)},t_{ki},g_{i}^{(k)}\right)\right\} _{k\in\partial i}\right)\nonumber \\
 & \propto & m_{i\to\psi_{i}}\left(t_{i}^{\left(j\right)},g_{i}^{\left(j\right)}\right)\sum_{t_{ki}}\prod_{k\in\partial i\setminus j}m_{k\to\psi_{i}}\left(t_{i}^{\left(j\right)},t_{ki},g_{i}^{\left(j\right)}\right)\times\\
 &  & \times\left[\delta\left(t_{i}^{(j)},0\right)+\delta\left(t_{i}^{(j)},\left(1+\min_{k\in\partial i}\left\{ t_{ki}\right\} \right)\right)\right]\nonumber \\
 & \propto & \delta\left(t_{i}^{\left(j\right)},0\right)m_{i\to\psi_{i}}\left(0,g_{i}^{\left(j\right)}\right)\prod_{k\in\partial i\setminus j}\sum_{t_{ki}}m_{k\to\psi_{i}}\left(0,t_{ki},g_{i}^{\left(j\right)}\right)+\label{eq:pass}\\
 &  & +\;m_{i\to\psi_{i}}\left(t_{i}^{\left(j\right)},g_{i}^{\left(j\right)}\right)\I\left(t_{i}^{\left(j\right)}\leq t_{ji}+1\right)\prod_{k\in\partial i\setminus j}\sum_{t_{ki}\geq t_{i}^{\left(j\right)}-1}m_{k\to\psi_{i}}\left(t_{i}^{\left(j\right)},t_{ki},g_{i}^{\left(j\right)}\right)\nonumber \\
 &  & -\;m_{i\to\psi_{i}}\left(t_{i}^{\left(j\right)},g_{i}^{\left(j\right)}\right)\I\left(t_{i}^{\left(j\right)}<t_{ji}+1\right)\prod_{k\in\partial i\setminus j}\sum_{t_{ki}>t_{i}^{\left(j\right)}-1}m_{k\to\psi_{i}}\left(t_{i}^{\left(j\right)},t_{ki},g_{i}^{\left(j\right)}\right)\nonumber 
\end{eqnarray}
where in (\ref{eq:pass}) we use the fact that 
\[
\delta\left(t_{i},\left(1+\min_{j\in\partial i}\left\{ t_{ji}\right\} \right)\right)=\prod_{j\in\partial i}\I\left(t_{i}\leq t_{ji}+1\right)-\prod_{j\in\partial i}\I\left(t_{i}<t_{ji}+1\right).
\]
Up to now, messages depend on the $T^{2}G$ values $\left(t_{i}^{\left(k\right)},t_{ki},g_{i}^{\left(k\right)}\right)$.
It is however possible to use more concise representation, retaining
just information on the relative timing between infection time~$t_{i}^{\left(j\right)}$
for a node~$i$ and the infection propagation time~$t_{ji}$ on
its link with node~$j$, introducing the variables

\begin{equation}
\sigma_{ji}=1+\sign\left(t_{ji}-\left(t_{i}^{\left(j\right)}-1\right)\right),\label{eq:sigmadef}
\end{equation}
In order to switch to the simplified representation with $\left(\sigma_{ji},\sigma_{ij}\right)$
variables defined in (\ref{eq:sigmadef}) instead of $t_{ji},t_{ij}$
ones, we will proceed as follows. In equation~(\ref{eq:pass}) we
can easily group the sums over different configurations of~$\left(t_{ki},t_{i}^{(j)}\right)$
and write: 
\begin{eqnarray}
p_{\psi_{i}\to j}\left(t_{i}^{\left(j\right)},\sigma_{ji},g_{i}^{\left(j\right)}\right) & \propto & \delta\left(t_{i}^{\left(j\right)},0\right)m_{i\to\psi_{i}}\left(0,g_{i}^{\left(j\right)}\right)\prod_{k\in\partial i\setminus j}\sum_{\sigma_{ki}}m_{k\to\psi_{i}}\left(0,\sigma_{ki},g_{i}^{\left(j\right)}\right)+\\
 &  & +\;m_{i\to\psi_{i}}\left(t_{i}^{\left(j\right)},g_{i}^{\left(j\right)}\right)\I\left(\sigma_{ji}=1,2\right)\prod_{k\in\partial i\setminus j}\sum_{\sigma_{ki}=1,2}m_{k\to\psi_{i}}\left(t_{i}^{\left(j\right)},\sigma_{ki},g_{i}^{\left(j\right)}\right)\nonumber \\
 &  & -\;m_{i\to\psi_{i}}\left(t_{i}^{\left(j\right)},g_{i}^{\left(j\right)}\right)\I\left(\sigma_{ji}=2\right)\prod_{k\in\partial i\setminus j}m_{k\to\psi_{i}}\left(t_{i}^{\left(j\right)},2,g_{i}^{\left(j\right)}\right)\nonumber 
\end{eqnarray}
Similarly, the outgoing message to the~$(t_{i},g_{i})$ variable
node is: 
\begin{eqnarray}
p_{\psi_{i}\to i}\left(t_{i},g_{i}\right) & \propto & \delta\left(t_{i},0\right)\prod_{k\in\partial i}\sum_{\sigma_{ki}}m_{k\to\psi_{i}}\left(0,\sigma_{ki},g_{i}\right)+\\
 &  & +\prod_{k\in\partial i}\sum_{\sigma_{ki}=1,2}m_{k\to\psi_{i}}\left(t_{i},\sigma_{ki},g_{i}\right)\nonumber \\
 &  & -\prod_{k\in\partial i}m_{k\to\psi_{i}}\left(t_{i},2,g_{i}\right)\nonumber 
\end{eqnarray}
In the simplified ~$\left(t,\sigma,g\right)$ representation for
the messages, the update equation for the~$\phi_{ij}$ nodes reads:
\begin{equation}
p_{\phi_{ij}\to j}\left(t_{j},\sigma_{ij},g_{j}\right)\propto\sum_{t_{i},\sigma_{ji},g_{i}}\Omega\left(t_{i},t_{j},\sigma_{ij},\sigma_{ji},g_{i},g_{j}\right)m_{i\to\phi_{ij}}\left(t_{i},\sigma_{ji},g_{i}\right)\label{eq:simplifiedphi}
\end{equation}
where: 
\begin{equation}
\Omega\left(t_{i},t_{j},\sigma_{ij},\sigma_{ji},g_{i},g_{j}\right)=\begin{cases}
\chi\left(t_{i},t_{j},\sigma_{ij},g_{i}\right) & :t_{i}<t_{j},~\sigma_{ji}=2,~\sigma_{ij}\neq2\\
\chi\left(t_{i},t_{j},\sigma_{ij},g_{i}\right)+\left(1-\lambda\right)^{g_{i}+1} & :t_{i}<t_{j},~\sigma_{ji}=2,~\sigma_{ji}=2\\
\chi\left(t_{j},t_{i},\sigma_{ji},g_{j}\right) & :t_{j}<t_{i},~\sigma_{ji}=2,~\sigma_{ji}\neq2\\
\chi\left(t_{j},t_{i},\sigma_{ji},g_{j}\right)+\left(1-\lambda\right)^{g_{j}+1} & :t_{j}<t_{i},~\sigma_{ij}=2,~\sigma_{ji}=2\\
1 & :t_{i}=t_{j},~\sigma_{ji}=\sigma_{ij}=2\\
0 & :otherwise
\end{cases}\label{eq:omega_phi_sigma}
\end{equation}
and 
\begin{equation}
\chi\left(t_{1},t_{2},\sigma,g\right)=\sum_{t=t_{1}}^{t_{1}+g}\delta\left(\sigma\left(t_{2},t\right),\sigma\right)\lambda\left(1-\lambda\right)^{t-t_{1}}\label{eq:chidef}
\end{equation}
Simple algebra and precalculation of terms in (\ref{eq:simplifiedphi})-(\ref{eq:chidef})
brings a significant optimization for updates involving the factor
node~$\phi_{ij}$ down to $O(TG^{2})$ operations per update.

\begin{center}
\rule[15pt]{1\textwidth}{0.5pt}
\par\end{center}

\section{Gradient descent updates}

The log-likelihood of the epidemic parameters is nothing but the free
energy of the model. In the Bethe approximation, it can be expressed
as a sum of local terms which only depends on the BP messages: 
\begin{equation}
-f=\sum_{a}f_{a}+\sum_{i}f_{i}-\sum_{(ia)}f_{(ia)}\label{eq:free-energy}
\end{equation}
where 
\begin{eqnarray}
f_{a} & = & \log\left(\sum_{\left\{ z_{i}:i\in\dd a\right\} }F_{a}\left(\left\{ z_{i}\right\} _{i\in\dd a}\right)\prod_{i\in\dd a}m_{i\to a}(z_{i})\right)\\
f_{(ia)} & = & \log\left(\sum_{z_{i}}m_{i\to a}(z_{i})p_{F_{a}\to i}(z_{i})\right)\\
f_{i} & = & \log\left(\sum_{z_{i}}\prod_{b\in\dd i}p_{F_{b}\to i}(z_{i})\right)
\end{eqnarray}
Since $f$ is a function of all the BP messages, one would argue that
this messages depend on the model parameters too, at every step in
the BP algorithm. Actually, there is no need to consider this implicit
$\left\{ \lambda_{ij},\mu_{i}\right\} $ dependence if BP has reached
its fixed point, that is when BP equations are satisfied and the messages
are nothing else but Lagrange multipliers with respect to the constraint
minimization of the Bethe free energy functional \citep{yedidia_bethe_2001}.
In the present parametrization, the only explicit dependence of free
energy on epidemic parameters is in the factor node terms $f_{a}$'s
involving the compatibility functions $\phi_{ij}=\omega_{ij}\left(t_{ij}-t_{i}|g_{i}\right)\omega_{ji}\left(t_{ji}-t_{j}|g_{j}\right)$
and $\mathcal{G}_{i}\left(g_{i}\right)=\mu_{i}\left(1-\mu_{i}\right)^{g_{i}}$,
and the gradient can be computed very easily. Please note that formulas
below show the derivative of the free energy $f=-\mathcal{L}$: the
GA updates of the log-likelihood only differ up to a minus sign. For
the $\phi_{ij}$ nodes we have: 
\begin{equation}
\frac{\partial f_{\phi_{ij}}}{\partial\lambda_{ij}}=\frac{\sum_{t_{i},t_{ji},g_{i},t_{j},t_{ij},g_{j}}\frac{\partial\phi_{ij}}{\partial\lambda_{ij}}\left(t_{i},t_{ji},g_{i},t_{j},t_{ij},g_{j}\right)m_{i\to\phi_{ij}}\left(t_{i},t_{ji},g_{i}\right)m_{j\to\phi_{ij}}\left(t_{j},t_{ij},g_{j}\right)}{\sum_{t_{i},t_{ji},g_{i},t_{j},t_{ij},g_{j}}\phi_{ij}\left(t_{i},t_{ji},g_{i},t_{j},t_{ij},g_{j}\right)m_{i\to\phi_{ij}}\left(t_{i},t_{ji},g_{i}\right)m_{j\to\phi_{ij}}\left(t_{j},t_{ij},g_{j}\right)}
\end{equation}
where 
\begin{equation}
\frac{\partial\phi_{ij}}{\partial\lambda_{ij}}=\begin{cases}
1 & t_{i}<t_{j}\mbox{ and }t_{i}=t_{ij}<t_{i}+g_{i}\\
-\left(g_{i}-t_{i}\right)\lambda_{ij}\left(1-\lambda_{ij}\right)^{g_{i}-t_{i}-1} & t_{i}<t_{j}\mbox{ and }t_{i}<t_{ij}=t_{i}+g_{i}\\
\left(1-\lambda_{ij}\right)^{t_{ij}-t_{i}}-\left(t_{ij}-t_{i}\right)\lambda_{ij}\left(1-\lambda_{ij}\right)^{t_{ij}-t_{i}-1} & t_{i}<t_{j}\mbox{ and }t_{i}<t_{ij}<t_{i}+g_{i}\\
1 & t_{j}<t_{i}\mbox{ and }t_{j}=t_{j}<t_{j}+g_{j}\\
-\left(g_{j}-t_{j}\right)\lambda_{ij}\left(1-\lambda_{ij}\right)^{g_{j}-t_{j}-1} & t_{j}<t_{i}\mbox{ and }t_{j}<t_{ji}=t_{j}+g_{j}\\
\left(1-\lambda_{ij}\right)^{t_{ji}-t_{j}}-\left(t_{ji}-t_{j}\right)\lambda_{ij}\left(1-\lambda_{ij}\right)^{t_{ji}-t_{j}-1} & t_{j}<t_{i}\mbox{ and }t_{j}<t_{ji}<t_{j}+g_{j}\\
0 & \mbox{else}
\end{cases}\label{eq:phi_ij_free_energy_tprime}
\end{equation}
In the simplified~$\left(t,\sigma,g\right)$ representation for the
messages, equation~(\ref{eq:phi_ij_free_energy_tprime}) takes the
form: 
\begin{equation}
\frac{\partial\phi_{ij}}{\partial\lambda_{ij}}=\begin{cases}
\chi\left(t_{i},t_{j},\sigma_{ij},g_{i}\right) & t_{i}<t_{j},\sigma_{ji}=2,\sigma_{ij}\neq2\\
\chi\left(t_{i},t_{j},\sigma_{ij},g_{i}\right)-\left(g_{i}+1\right)\left(1-\lambda\right)^{g_{i}} & t_{i}<t_{j},\sigma_{ji}=2,\sigma_{ij}=2\\
\chi\left(t_{j},t_{i},\sigma_{ji},g_{j}\right) & t_{j}<t_{i},\sigma_{ji}=2,\sigma_{ij}\neq2\\
\chi\left(t_{j},t_{i},\sigma_{ji},g_{j}\right)-\left(g_{j}+1\right)\left(1-\lambda\right)^{g_{j}} & t_{j}<t_{i},\sigma_{ji}=2,\sigma_{ij}=2\\
0 & otherwise
\end{cases}\label{eq:phi_ij_free_energy_sigma}
\end{equation}
where: 
\begin{equation}
\chi\left(t_{1},t_{2},\sigma,g\right)=\sum_{t=t_{1}}^{t_{1}+g}\delta\left(\sigma\left(t_{2},t\right),\sigma\right)\left(1-\lambda_{ij}\right)^{t-t_{1}}-\left(t-t_{1}\right)\lambda_{ij}\left(1-\lambda_{ij}\right)^{t-t_{1}-1}
\end{equation}
For the $\mathcal{G}_{i}$ nodes we have: 
\begin{equation}
\frac{\partial f_{\mathcal{G}_{i}}}{\partial\mu_{i}}=\frac{\sum_{g_{i}}\mathcal{\tilde{G}}_{i}(g_{i})m_{i\to\mathcal{G}_{i}}(g_{i})}{\sum_{g_{i}}\mathcal{G}_{i}(g_{i})m_{i\to\mathcal{G}_{i}}(g_{i})}
\end{equation}
where 
\begin{equation}
\mathcal{\tilde{G}}_{i}(g_{i})=\begin{cases}
\left(1-\mu_{i}\right)^{g_{i}}-g_{i}\mu_{i}\left(1-\mu_{i}\right)^{g_{i}-1} & :g_{i}<G\\
G-G\left(1-\mu_{i}\right)^{G-1} & :g_{i}=G.
\end{cases}
\end{equation}

\section*{Acknowledgments}

We warmly thank L. Dall'Asta for useful discussions, and Riccardo
Refolo for providing us with Fig. 1. AB and APM acknowledge support
by Fondazione CRT, project SIBYLunder the initiative \textquotedblleft La
Ricerca dei Talenti\textquotedblright . 

\section*{Author Contributions}

AB, AI and APM contributed equally to this work.

\noindent \bibliographystyle{unsrtnat}
\bibliography{references}

\end{document}